\documentclass[%
unsortedaddress,
 amsmath,amssymb,amsfonts,
 aps,
]{revtex4-2}

\usepackage[dvipdfmx]{graphicx}
\usepackage{dcolumn}
\usepackage{bm}
\usepackage{xcolor}

\newcommand{\tabitem}{~~\llap{\textbullet}~~}

\def\ket#1{\mbox{\boldmath $#1$}}

\newcommand{\enter}{\curvearrowleft}
\newcommand{\inter}{\curvearrowright}
\newcommand{\intra}{\circlearrowright}
\newcommand{\Eq}{equation}
\newcommand{\stMapEqn}{single-trajectory map equation}
\newcommand{\ACL}{average code length}
\newcommand{\Methods}{Methods section}
\newcommand{\Supplemental}{Supplementary Information}
\newcommand{\Supplementary}{Supplementary}
\newcommand{\Sec}{Section}
\newcommand{\SBM}{SBM}

\begin{document}


\title{Single-trajectory map equation}


\author{Tatsuro Kawamoto}
\affiliation{Artificial Intelligence Research Center, \\
  National Institute of Advanced Industrial Science and Technology, 
  Tokyo, Japan }


\date{\today}

\begin{abstract}
Community detection, the process of identifying module structures in complex systems represented on networks, is an effective tool in various fields of science. The map equation, which is an information-theoretic framework based on the random walk on a network, is a particularly popular community detection method. 
Despite its outstanding performance in many applications, the inner workings of the map equation have not been thoroughly studied. 
Herein, we revisit the original formulation of the map equation and address the existence of its ``raw form,'' which we refer to as the single-trajectory map equation. 
This raw form sheds light on many details behind the principle of the map equation that are hidden in the steady-state limit of the random walk. 
Most importantly, the single-trajectory map equation provides a more balanced community structure, naturally reducing the tendency of the overfitting phenomenon in the map equation.
\end{abstract}


\maketitle

\section*{Introduction}
Community detection plays a vital role in various disciplines of science dealing with network data. 
It is an unsupervised learning task to classify the node set in a network into groups, or modules. 
Each subgraph that is identified as a module has no significant internal structure, while we expect each module to be structurally distinct. 
Thus, community detection provides a concise explanation of the dataset by ignoring detailed structures as noises within a module while preserving a significant macroscopic organisation as a module structure \cite{SCHAEFFER2007,Fortunato2010,Fortunato2016,Jin2021}. 
Analogous to other machine-learning tasks, because it is often not apparent which parts of the dataset represent noise, community detection algorithms suffer from overfitting and underfitting problems \cite{Ghasemian2020}. 

\begin{figure}[t!]
  \centering
  \includegraphics[width= 0.6\columnwidth]{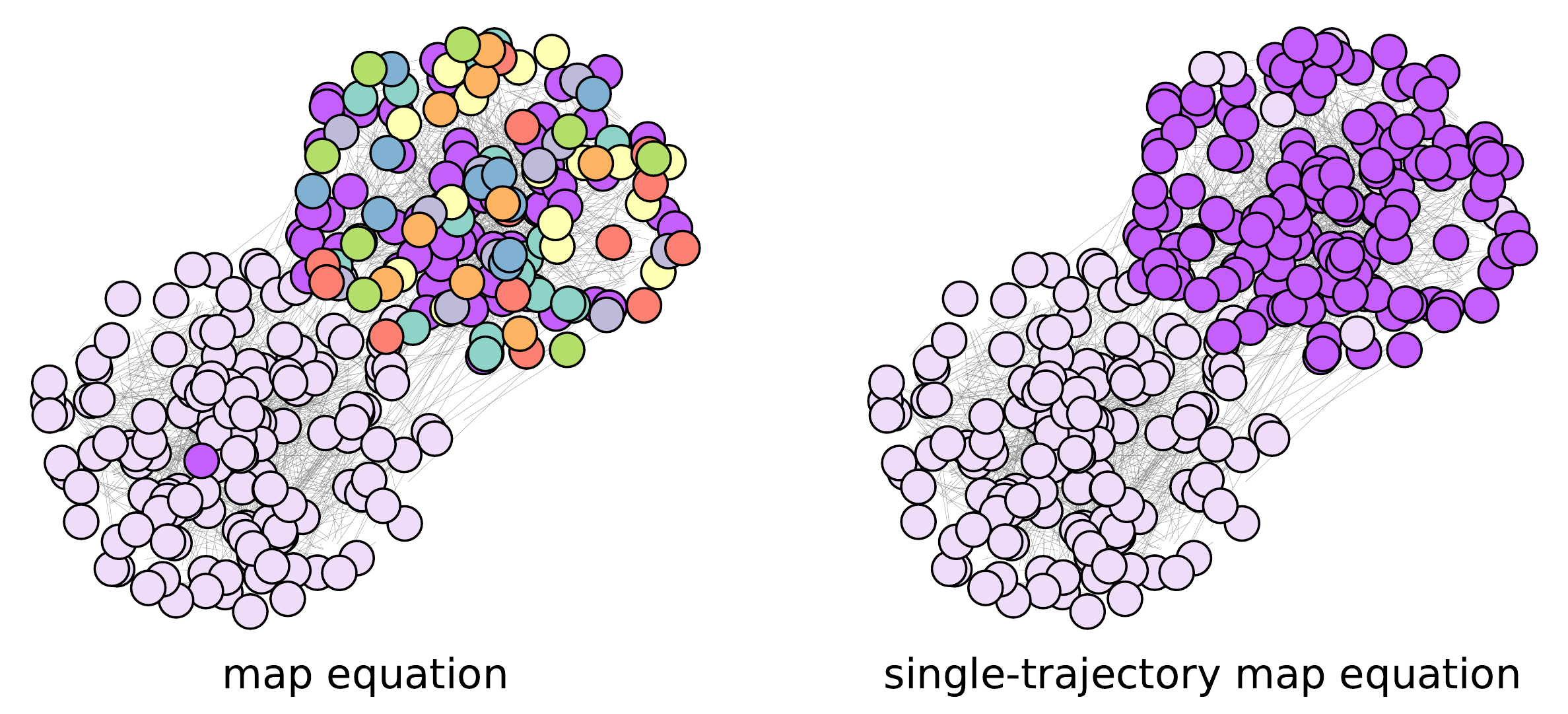}
  \caption{
  Community detection based on the map equation and the {\stMapEqn} applied to a synthetic network.
  The nodes in the same colour belong to the same module. 
  See Experiments section for details. 
}
  \label{fig:SBM}
\end{figure}

The map equation \cite{Rosvall2008} is a popular community detection method for networks and is formulated as a minimization problem of an information-theoretic objective function describing the average code length of the random walk. 
Infomap \cite{MapEquationURL}---an implementation for a greedy optimisation of the map equation---has often been used to analyze real-world datasets. 
Furthermore, there are several extensions of the map equation itself \cite{Rosvall2011,Esquivel2011,Rosvall2014,DeDomenico2015,Kheirkhahzadeh2016,edler2017mapping,Aslak2018,Emmons2019,smiljanic2019mapping,Blocker2020,eriksson2021choosing,smiljanic2021}, mainly focusing on incorporating higher-order network information. 
However, the map equation is prone to overfitting, particularly for sparse networks. 
Despite the map equation providing an optimal module structure for the description of the random walk on a network, an excessively fine module structure may be obtained. 
For example, as illustrated in Fig.~\ref{fig:SBM} (left), many small modules are often identified in addition to a few large modules. 
When too many modules consisting of only a few nodes are identified, we have difficulty interpreting it as a concise explanation of the dataset. 

This study revisits the map equation and considers its raw form, which we refer to as the \textit{single-trajectory} map equation. 
Its objective function is the average code length of (not necessarily random) walkers with finite path lengths. 
The concept of the {\stMapEqn} already appears in the original paper \cite{Rosvall2008} for a schematic description of the map equation. 
Nevertheless, this raw form has never been actively studied or utilised although it is a valuable variant with a mechanism to prune small modules and prevent the map equation from overfitting (as depicted in Fig.~\ref{fig:SBM} (right)).

The emergence of small or highly unbalanced modules has been discussed in various contexts in the community detection literature. 
It is often considered to be the nature of real-world datasets \cite{Arenas2004,Clauset2004,Leskovec2009}, or a phenomenon that occurs because of the implementation details of an algorithm \cite{Clauset2004,WakitaTsurumi2007}. 
In the context of the inference problem, the emergence of small modules is interpreted as artefacts caused by overfitting. 
Optimization-based (or maximum likelihood-based) methods are typically prone to overfitting, whereas methods based on Bayesian formulations avoid partitioning a network or subgraph where there is no statistically significant internal structure \cite{MooreReview2017,Peixoto2017tutorial}, that is, they avoid generating small modules. 
The map equation also has a Bayesian counterpart \cite{smiljanic2019mapping,smiljanic2021}. 
Regardless of the underlying mechanism, the pruning of small modules is sometimes preferred in practice because it provides a more concise explanation of the network. 
A similar issue can be found in the regression problem in supervised learning \cite{HastieTibshiraniFriedman}. 
Although the ridge regression is a principled method, many variables with extremely small coefficients are often assessed as significant. 
In contrast, the lasso regression prunes such variables and provides a concise description of the dataset.

The {\stMapEqn} is a variant of the map equation that achieves a partition of coarser-resolution scale. 
Our approach differs from that of the hierarchical map equation \cite{Rosvall2011} that achieves finer-resolution partitions \cite{KawamotoRosvall2015}. 
Although a high-resolution community detection method is useful when the network consists of several small modules, a method with a coarser resolution is also needed when an algorithm suffers from overfitting. 
For bipartite networks, \cite{Blocker2020} showed that a coarser resolution can be obtained by incorporating the bipartiteness property. 
A different resolution scale can also be obtained by introducing the ``Markov time'' \cite{Kheirkhahzadeh2016,Schaub2012}, which is an external parameter of the random walk. 
However, as shown in the following, the framework of the map equation can intrinsically prune small modules when formulated as the {\stMapEqn} and the balanced-size modules can be identified in a principled manner.

\section*{Results}

\subsection*{Revisiting the map equation}
\begin{figure*}[t!]
  \centering
  \includegraphics[width= 0.95 \columnwidth]{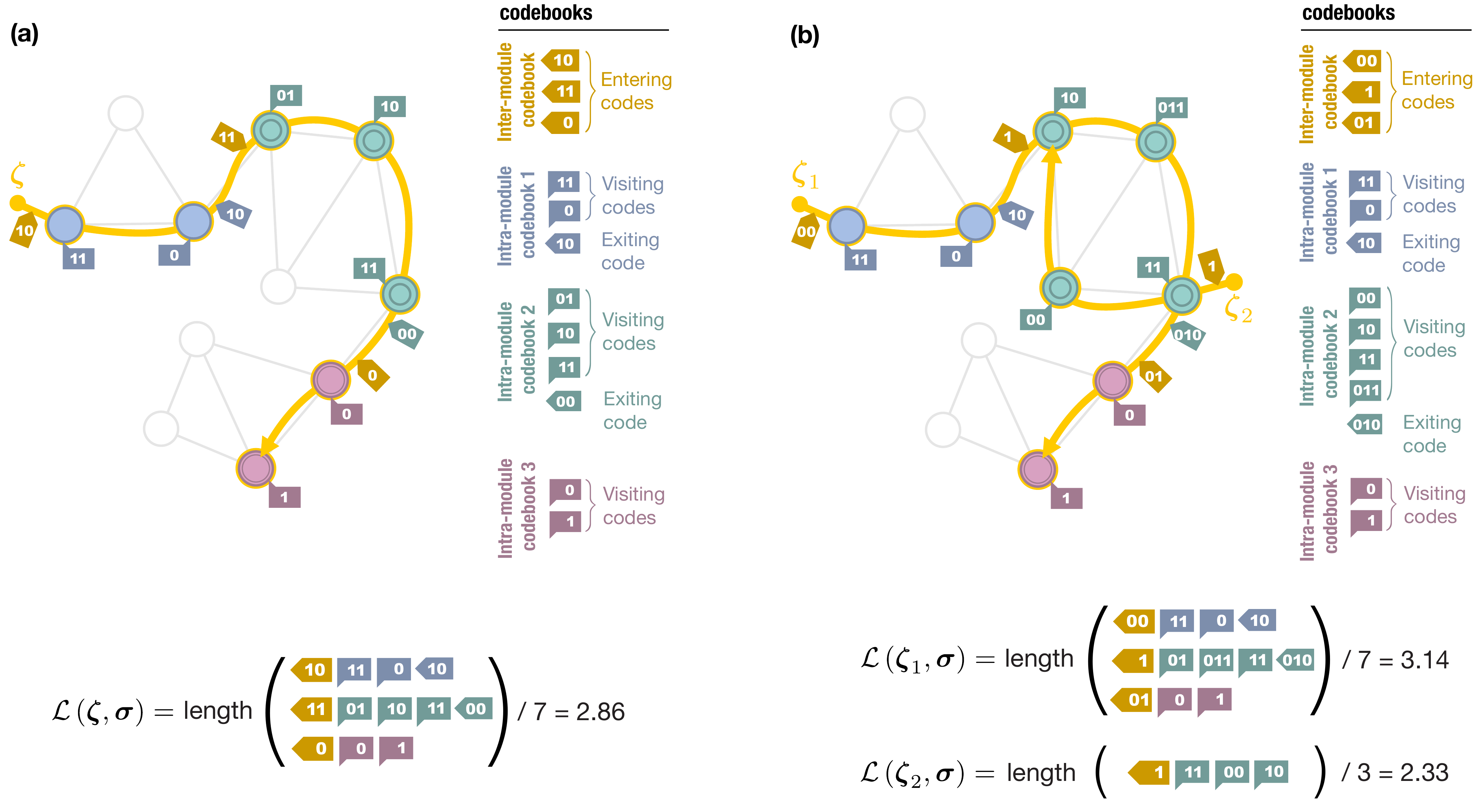}
  \caption{
Trajectories (yellow solid lines) on a network and their encoding for a given node partition. 
Nodes in the same module have the same symbol and colour representations. 
The codewords in each codebook are listed on the right. 
Whereas there is only one trajectory in {\bf (a)}, another trajectory is added in {\bf (b)}. 
The {\ACL} of each trajectory is shown at the bottom. 
}
  \label{fig:stMapEqnExample}
\end{figure*}

We proceed with the step-by-step formulation of the map equation. 
A prominent characteristic of the map equation is the hierarchical encoding scheme for the random walk using multiple codebooks that takes account of module structure in a network. 
As a specific example, let us consider the encoding of a trajectory in a network as shown in Fig.~\ref{fig:stMapEqnExample}{\bf (a)}. 
We let $\ket{\zeta} = \{ \zeta_{0}, \dots, \zeta_{T-1} \}$ be a trajectory of a walker, where $\zeta_{t} \in V$ is the $t$th visited node. 
We also consider a partition $\ket{\sigma} = \{ \sigma_{1}, \dots, \sigma_{N} \}$ of node set $V$ ($|V| = N$), where $\sigma_{i} \in \{1, \dots, K\}$ is the module label of node $i \in V$. 
$K$ is the number of modules. 
A trajectory of a walker is encoded using two types of codebooks: inter-module and intra-module codebooks. 
The inter-module codebook describes the transitions of a walker moving into another module. 
In contrast, each intra-module codebook describes the walker transiting between nodes within the module or exiting the module.

The actual codewords for a trajectory based on the Huffman coding \cite{Mackay2003,cover1999elements} are shown in Fig.~\ref{fig:stMapEqnExample}{\bf (a)}. 
Starting with the code ``\textsf{10}'' for the module that the walker has first visited, the trajectory is described by indicating the visited nodes. 
Every time the walker moves to a different module, the exiting code of the previous module and the entering code of the next module are consumed. 

In general, given a trajectory $\ket{\zeta}$ of length $T$ and module assignments $\ket{\sigma}$, the average code length $\mathcal{L}\left( \ket{\zeta}, \ket{\sigma} \right)$ is expressed as 
\begin{align}
&\mathcal{L}\left( \ket{\zeta}, \ket{\sigma} \right) 
= 
\frac{1}{T} \Biggl( 
\sum_{t=0}^{T-1} \ell_{0}\left( \zeta_{t}, \sigma_{\zeta_{t}} \right) 
+ \sum_{\substack{t=0 \\ (\sigma_{\zeta_{t}} \ne \sigma_{\zeta_{t+1}})} }^{T-2} 
\biggl( \ell_{0}\left(\inter, \sigma_{\zeta_{t}} \right) 
+ \ell_{1}\left( \sigma_{\zeta_{t+1}} \right) \biggr)
+\ell_{1}\left( \sigma_{\zeta_{0}} \right) 
\Biggr). \label{stMapEqn1}
\end{align}
Here, we denote $\ell_{0}\left(i, \sigma \right)$ as the length of the code in an intra-module codebook, indicating that a walker visits node $i \in V$ in module $\sigma$; 
$\ell_{0}\left(\inter, \sigma \right)$ as the length of the code in an intra-module codebook, indicating that a walker exits module $\sigma$; and 
$\ell_{1}\left(\sigma \right)$ as the length of the code in an inter-module codebook, indicating that a walker enters module $\sigma$. 
In {\Eq}~(\ref{stMapEqn1}), the first summation represents the code length for visited nodes and the second summation represents the code length for transitions between modules. 
The last term is the code length for the module at the starting point, with a negligible contribution when $T \gg 1$. 
It is important that the codebooks are coupled; that is, because an exiting code from a module belongs to an intra-module codebook, transitions between modules affect the encoding of transitions within each module. 

The principle of the map equation framework is that the compression of the {\ACL} through the hierarchical coding reveals a module structure as an optimal partition $\ket{\sigma}$. 
Readers might believe that the introduction of codewords for transitions between modules simply makes the code length longer. 
However, such a hierarchical encoding scheme can compress the {\ACL} because it allows us to assign shorter codewords for visited nodes; for example, although the code ``\textsf{0}'' is assigned to two different nodes in Fig.~\ref{fig:stMapEqnExample}{\bf (a)}, they are distinguishable because they belong to different modules. 
Therefore, when a trajectory rarely consumes the codewords for transitions between modules, the {\ACL} can be compressed more efficiently.

Equation (\ref{stMapEqn1}) can also be expressed using visiting frequencies as follows: 
\begin{align}
&\mathcal{L}\left( \ket{\zeta}, \ket{\sigma} \right) 
= \left( \frac{1}{T} + \sum_{\tilde{\sigma}^{\prime}, \tilde{\sigma}} \hat{\mathsf{p}}_{ \tilde{\sigma}^{\prime} \tilde{\sigma} } \right) \mathcal{H}_{1} 
+ \sum_{\sigma} \left( \sum_{j \in \sigma} \hat{\mathsf{p}}_{j} + \sum_{\tilde{\sigma}^{\prime}} \hat{\mathsf{p}}_{ \tilde{\sigma}^{\prime} \sigma } \right) \mathcal{H}^{\sigma}_{0}, \label{stMapEqn2} \\
&\begin{cases}
\mathcal{H}_{1} 
= \sum_{\sigma^{\prime}} \frac{\sum_{\sigma}\hat{\mathsf{p}}_{ \sigma^{\prime} \sigma }}{ 1/T + \sum_{\tilde{\sigma}^{\prime}, \tilde{\sigma}} \hat{\mathsf{p}}_{ \tilde{\sigma}^{\prime} \tilde{\sigma} } } \ell_{1}\left( \sigma^{\prime} \right) + \frac{ 1/T }{ 1/T + \sum_{\tilde{\sigma}^{\prime}, \tilde{\sigma}} \hat{\mathsf{p}}_{\tilde{\sigma}^{\prime} \tilde{\sigma} } } \ell_{1}\left( \sigma_{\zeta_{0}} \right) \\
\mathcal{H}^{\sigma}_{0} = \sum_{i \in \sigma} \frac{\hat{\mathsf{p}}_{i}}{ \sum_{j \in \sigma} \hat{\mathsf{p}}_{j} + \sum_{\tilde{\sigma}^{\prime}} \hat{\mathsf{p}}_{ \tilde{\sigma}^{\prime} \sigma } } \ell_{0}\left( i, \sigma \right) 
+ \frac{ \sum_{\sigma^{\prime}} \hat{\mathsf{p}}_{ \sigma^{\prime} \sigma } }{ \sum_{j \in \sigma} \hat{\mathsf{p}}_{j} + \sum_{\tilde{\sigma}^{\prime}} \hat{\mathsf{p}}_{ \tilde{\sigma}^{\prime} \sigma } } \ell_{0}\left( \inter, \sigma \right) 
\end{cases}, \label{ConditionalACL}
\end{align}
where $\sum_{i \in \sigma}$ represents the sum over the node set in module $\sigma$. 
$\hat{\mathsf{p}}_{i}$ is the visiting frequency of node $i \in V$ and $\hat{\mathsf{p}}_{ \sigma^{\prime} \sigma}$ is the joint transition frequency from module $\sigma$ to module $\sigma^{\prime}$, i.e., 
\begin{align}
& \hat{\mathsf{p}}_{i} = \frac{1}{T} \sum_{t=0}^{T-1} \delta_{i, \zeta_{t}}, \\
& \hat{\mathsf{p}}_{\sigma^{\prime} \sigma} = 
\begin{cases}
\frac{1}{T} \sum_{t=0}^{T-2} \delta_{\sigma^{\prime}, \sigma_{\zeta_{t+1}}} \delta_{\sigma, \sigma_{\zeta_{t}}} & (\text{for } \sigma \ne \sigma^{\prime})\\
0  & (\text{for } \sigma = \sigma^{\prime})
\end{cases}, \label{pNodepModule}
\end{align}
where $\delta_{ab}$ represents the Kronecker delta. 
$\mathcal{H}^{\sigma}_{0}$ and $\mathcal{H}_{1}$ are conditional average code lengths within the intra- and inter-module codebooks, respectively.

Recall that the random walk is a stochastic variable; there is no such thing as a single (finite-length) trajectory representing the random walk. 
Therefore, instead of a specific trajectory, we consider the \textit{expected} average code length $\mathbb{E}_{\scriptsize \ket{Z}}\left[\mathcal{L}\left( \ket{Z}, \ket{\sigma} \right) \right]$ in the map equation, where $\ket{Z}$ is the stochastic variable representing the random walk; in other words, $\mathbb{E}_{\scriptsize \ket{Z}}[\cdots]$ is the ensemble average over all possible trajectories (say, from all possible starting points). 
We assume that the trajectory length $T$ is sufficiently large that the random walk is in a steady state. 
When the network is strongly connected, the empirical frequencies are converted to the corresponding steady-state probabilities. 
$\sum_{\sigma} \hat{\mathsf{p}}_{\sigma^{\prime} \sigma}$ is converted to the entering probability into module $\sigma^{\prime}$ denoted by $q_{\sigma^{\prime} \enter}$; 
$\sum_{\sigma^{\prime}} \hat{\mathsf{p}}_{\sigma^{\prime} \sigma}$ to the exiting probability from module $\sigma$ denoted by $q_{\sigma \inter}$; and 
$\hat{\mathsf{p}}_{i}$ to the visiting probability of node $i$ denoted by $q_{i}$. 
The conditional average code lengths $\mathcal{H}^{\sigma}_{0}$ and $\mathcal{H}_{1}$ are also converted to the expectations. 
According to Shannon's source coding theorem \cite{cover1999elements,shannon1948mathematical}, these expectations are respectively bounded by the Shannon entropies, 
\begin{align}
H_{1}\left( \{ q_{\sigma \enter} \} \right) &= -\sum_{\sigma=1}^{K} \frac{q_{\sigma \enter}}{q_{\enter}} \log \frac{q_{\sigma \enter}}{q_{\enter}}, \\
H^{\sigma}_{0}\left( q_{\sigma \inter}, \{ q_{i} \}_{i\in \sigma} \right) &= 
-\frac{q_{\sigma \inter}}{p^{\sigma}_{\intra}} \log \frac{q_{\sigma \inter}}{p^{\sigma}_{\intra}} 
-\sum_{i \in \sigma} \frac{q_{i}}{p^{\sigma}_{\intra}} \log \frac{q_{i}}{p^{\sigma}_{\intra}},
\end{align}
where $q_{\enter} = \sum_{\sigma=1}^{K} q_{\sigma \enter}$, $p^{\sigma}_{\intra} = q_{\sigma \inter} + \sum_{i \in \sigma} q_{i}$, and $\log$ is the logarithm with base 2. 
Then, the expected average code length of the random walk is bounded from below as follows: 
\begin{align}
L\left( \ket{\sigma} \right) 
&= q_{\enter} H_{1}\left( \{ q_{\sigma \enter} \} \right) + \sum_{\sigma} p^{\sigma}_{\intra} H^{\sigma}_{0}\left( q_{\sigma \inter}, \{ q_{i} \}_{i\in \sigma} \right) 
\le \mathbb{E}_{\scriptsize \ket{Z}}\left[\mathcal{L}\left( \ket{Z}, \ket{\sigma} \right) \right]. \label{MapEqn}
\end{align}
Note here that the contribution from the starting points of the random walk is excluded. 
This lower bound asymptotically coincides with the expected average code length itself as $T \to\infty$. 
This is the objective function of the map equation and the node partition $\ket{\sigma}$ is optimised so that $L\left( \ket{\sigma} \right)$ is minimised.

The assumption that the network is strongly connected plays a vital role in the aforementioned derivation. 
If this is not the case, the trajectory length $T$ cannot be sufficiently large. 
Then, the contribution from the starting points of the random walk may not be negligible in $\mathbb{E}_{\scriptsize \ket{Z}}\left[\mathcal{L}\left( \ket{Z}, \ket{\sigma} \right) \right]$. 
Therefore, we can say that the map equation evaluates the code length of the ``flow.'' 
It is a stochastic variable representing the ensemble of transitions, and it has no information about the starting points of the random walk by definition (as discussed below, this distinction becomes more prominent when we consider the \texttt{-\,-flow-model rawdir} option in Infomap). 
The only input for the map equation is a network because the connectivity of nodes fully characterises the flow. 
By the introduction of so-called teleportation \cite{brin1998anatomy} to the random walk that moves the walker to another node randomly with a certain probability, we can always let the trajectory length $T$ be infinitely large and make the flow ergodic \cite{Rosvall2008}. 
Therefore, the map equation is not essentially limited to strongly-connected networks.

\subsection*{Single-trajectory map equation}
The average code length $\mathcal{L}\left( \ket{\zeta}, \ket{\sigma} \right)$ of a trajectory is the raw form of the objective function in the map equation. 
When we have multiple trajectories $\{\ket{\zeta}_{a}\} := \{\ket{\zeta}_{1}, \dots, \ket{\zeta}_{M}\}$ on a common node set, analogous to the expected average code length $\mathbb{E}_{\scriptsize \ket{Z}}\left[\mathcal{L}\left( \ket{Z}, \ket{\sigma} \right) \right]$, we consider the following mean {\ACL}: 
\begin{align}
\overline{\mathcal{L}}\left( \ket{\sigma}; \{\ket{\zeta}_{a}\} \right) 
= \frac{1}{M} \sum_{a=1}^{M} \mathcal{L}\left( \ket{\zeta}_{a}, \ket{\sigma} \right). \label{AveragedStMapEqn}
\end{align}
Each trajectory may have different lengths. 
Similar to the $L\left( \ket{\sigma} \right)$, this mean {\ACL} can be used as a minimization function to determine the optimal module assignments of nodes. 
We refer to such an optimisation method as the {\stMapEqn}. 
Note that the trajectories $\{\ket{\zeta}_{a}\}$ are provided as inputs in equation (\ref{AveragedStMapEqn}); unlike the map equation, there is no need to assume that they are generated (or simulated) from random walks, although one can consider simulated walks as trajectories. 

The average code lengths in the summation of {\Eq}~(\ref{AveragedStMapEqn}) are not independent because they share codebooks. 
To illustrate this, let us consider two trajectories as shown in Fig.~\ref{fig:stMapEqnExample}{\bf (b)}. 
Although the trajectory $\ket{\zeta}_{1}$ is identical to $\ket{\zeta}$ in Fig.~\ref{fig:stMapEqnExample}{\bf (a)}, the codes describing them are different because we must assign codewords for the nodes that $\ket{\zeta}_{1}$ does not go through due to the existence of trajectory $\ket{\zeta}_{2}$. 
In contrast, the nodes where no trajectories go through do not contribute to the average code lengths, reflecting the fact that the trajectories of finite lengths are considered. 
Those nodes should not have any module labels because there is no information based on the trajectories.

As we have seen, $L\left( \ket{\sigma} \right)$ and $\overline{\mathcal{L}}\left( \ket{\sigma}; \{\ket{\zeta}_{a}\} \right)$ are conceptually different. 
In the map equation, $L\left( \ket{\sigma} \right)$ is the expected average code length for the flow that is completely specified by the transition probabilities. 
We can also modify the transitions using teleportation to make the random walk ergodic. 
By contrast, $\overline{\mathcal{L}}\left( \ket{\sigma}; \{\ket{\zeta}_{a}\} \right)$ does not have such a stochasticity. 
It is the mean of the actual average code lengths, where each element corresponds to a single trajectory. 
Furthermore, $\overline{\mathcal{L}}\left( \ket{\sigma}; \{\ket{\zeta}_{a}\} \right)$ depends explicitly on the coding scheme applied, e.g., the Huffman coding, Shannon--Fano coding \cite{cover1999elements}, etc. 
Quantitatively, the contribution from the last term in {\Eq}~(\ref{stMapEqn1}) mainly makes the minimization of $\overline{\mathcal{L}}\left( \ket{\sigma}; \{\ket{\zeta}_{a}\} \right)$ distinct from that of $L\left( \ket{\sigma} \right)$. 
The codeword for the module that is required to specify the starting point of a trajectory makes the coding using multiple codebooks less efficient. Recall that an efficient compression is achieved when the inter-module codebook is not frequently used. 
This implies that the introduction of module labels is more costly in $\overline{\mathcal{L}}\left( \ket{\sigma}; \{\ket{\zeta}_{a}\} \right)$ and the {\stMapEqn} avoids generating many small modules.

The {\stMapEqn} searches for the node partition $\ket{\sigma}$ that achieves the optimal compression for the description of trajectories under a certain coding scheme. 
The optimality of the coding scheme itself is not required for the effectiveness of the method. 
Therefore, we can use different types of coding for the intra-module and inter-module codebooks. 
For example, we can introduce a heterogeneous coding where the code lengths are multiplied by a constant factor $\lambda > 0$ for the codewords in the inter-module codebook. 
That is, given a trajectory $\ket{\zeta}$ and a partition $\ket{\sigma}$, {\Eq}~(\ref{stMapEqn1}) is modified to 
\begin{align}
&\mathcal{L}_{\lambda}\left( \ket{\zeta}, \ket{\sigma} \right) 
= 
\frac{1}{T} \Biggl( 
\sum_{t=0}^{T-1} \ell_{0}\left( \zeta_{t}, \sigma_{\zeta_{t}} \right) 
+ \sum_{\substack{t=0 \\ (\sigma_{\zeta_{t}} \ne \sigma_{\zeta_{t+1}})} }^{T-2} 
\biggl( \ell_{0}\left(\inter, \sigma_{\zeta_{t}} \right) 
+ \lambda \ell_{1}\left( \sigma_{\zeta_{t+1}} \right) \biggr)
+\lambda \ell_{1}\left( \sigma_{\zeta_{0}} \right) 
\Biggr). \label{stMapEqn1Lambda}
\end{align}
Here, $\lambda$ is a hyperparameter that penalises the emergence of modules when such modules are relatively inefficient for the compression of the code length.

We can also derive a lower bound for the actual code length using Shannon's source coding theorem, similar to how $L\left( \ket{\sigma} \right)$ was such an estimate for the random walk in the steady-state limit. 
To this end, we consider the {\ACL} of the concatenated code, $\sum_{a=1}^{M} T_{a} \mathcal{L}\left( \ket{\zeta}_{a}, \ket{\sigma} \right)/\sum_{a=1}^{M} T_{a}$, where $T_{a}$ is the length of the $a$th trajectory; equivalently $\overline{\mathcal{L}}\left( \ket{\sigma}; \{\ket{\zeta}_{a}\} \right)$ when all trajectories have the same length. 
We regard the empirical frequencies $\hat{\mathsf{p}}_{i}$ and $\hat{\mathsf{p}}_{\sigma^{\prime} \sigma}$ as the true probabilities for the stochastic variables indicating the codewords and the concatenated code as the expected code length. 
Then, the conditional average code lengths in {\Eq}~(\ref{ConditionalACL}) are bounded from below by the Shannon entropies \cite{cover1999elements} with the empirical frequencies. 
Therefore, the {\ACL} of the concatenated code is bounded as follows:
\begin{align}
\underline{\mathcal{L}}\left( \ket{\sigma}; \{\ket{\zeta}_{a}\} \right) 
&= \hat{\mathsf{q}}_{\enter} H_{1}\left( \{ \hat{\mathsf{q}}_{\sigma \enter} \} \right) + \sum_{\sigma} \hat{\mathsf{p}}^{\sigma}_{\intra} H^{\sigma}_{0}\left( \hat{\mathsf{q}}_{\sigma \inter}, \{ \hat{\mathsf{q}}_{i} \}_{i\in \sigma} \right) 
\le \frac{1}{\sum_{a=1}^{M} T_{a}}\sum_{a=1}^{M} T_{a} \mathcal{L}\left( \ket{\zeta}_{a}, \ket{\sigma} \right), \label{stMapEqnConcat}
\end{align}
where 
\begin{align}
& \hat{\mathsf{q}}_{i} = \frac{1}{\sum_{a=1}^{M} T_{a}} \sum_{a=1}^{M} \sum_{t=0}^{T_{a}-1} \delta_{i, \zeta_{a t}}, \notag\\
& \hat{\mathsf{q}}_{\sigma^{\prime} \sigma} = 
\begin{cases}
\displaystyle \frac{1}{\sum_{a=1}^{M} T_{a}} \sum_{a=1}^{M} \sum_{t=0}^{T_{a}-2} 
\delta_{\sigma^{\prime}, \sigma_{\zeta_{a t+1}}} 
\delta_{\sigma, \sigma_{\zeta_{a t}}} 
& (\sigma \ne \sigma^{\prime})\\
0  & (\sigma = \sigma^{\prime})
\end{cases}, \notag\\
& \hat{\mathsf{q}}_{\sigma^{\prime} \enter} = \frac{ \sum_{a=1}^{M} \delta_{\sigma^{\prime}, \sigma_{\zeta_{a 0}} } }{\sum_{a=1}^{M} T_{a}}  
+ \sum_{\sigma} \hat{\mathsf{q}}_{\sigma^{\prime} \sigma}, \hspace{20pt} 
\hat{\mathsf{q}}_{\sigma \inter} = \sum_{\sigma^{\prime}} \hat{\mathsf{q}}_{\sigma^{\prime} \sigma}, \notag\\
& \hat{\mathsf{q}}_{\enter} = \sum_{\sigma} \hat{\mathsf{q}}_{\sigma \enter}, \hspace{20pt} 
\hat{\mathsf{p}}^{\sigma}_{\intra} = \hat{\mathsf{q}}_{\sigma \inter} + \sum_{i \in \sigma} \hat{\mathsf{q}}_{i}. \label{qInterIntraProbs}
\end{align}
We regard $\underline{\mathcal{L}}\left( \ket{\sigma}; \{\ket{\zeta}_{a}\} \right)$ as an alternative objective function for the {\stMapEqn}. 
$\underline{\mathcal{L}}\left( \ket{\sigma}; \{\ket{\zeta}_{a}\} \right)$ is independent of the coding scheme and its minimization is computationally more efficient than that of $\overline{\mathcal{L}}\left( \ket{\sigma}; \{\ket{\zeta}_{a}\} \right)$ because we do not need to construct the codebooks explicitly. 
Note that $\hat{\mathsf{q}}_{\sigma \enter}$ and $\hat{\mathsf{q}}_{\sigma \inter}$ may not coincide in {\Eq}~(\ref{qInterIntraProbs}), whereas $q_{\sigma \enter} = q_{\sigma \inter}$ for any module in $L\left( \ket{\sigma} \right)$ owing to the detailed balance condition of the random walk in the steady state. 
Analogous to $\mathcal{L}_{\lambda}\left( \ket{\zeta}, \ket{\sigma} \right)$ in {\Eq}~(\ref{stMapEqn1Lambda}), we can also consider a heterogeneous coding in $\underline{\mathcal{L}}\left( \ket{\sigma}; \{\ket{\zeta}_{a}\} \right)$, i.e., 
\begin{align}
\underline{\mathcal{L}}_{\lambda}\left( \ket{\sigma}; \{\ket{\zeta}_{a}\} \right) 
&= \lambda \, \hat{\mathsf{q}}_{\enter} H_{1}\left( \{ \hat{\mathsf{q}}_{\sigma \enter} \} \right) 
+ \sum_{\sigma} \hat{\mathsf{p}}^{\sigma}_{\intra} H^{\sigma}_{0}\left( \{ \hat{\mathsf{q}}_{\sigma \inter} \}, \{ \hat{\mathsf{q}}_{i} \}_{i\in \sigma} \right). \label{stMapEqnConcatLambda}
\end{align}

Interestingly, the method using $\underline{\mathcal{L}}\left( \ket{\sigma}; \{\ket{\zeta}_{a}\} \right)$ is also related to a variant of the map equation that is implemented in Infomap as an option named \texttt{-\,-flow-model rawdir}. 
In this variant of the map equation, we consider the \textit{flow based on the set of transition probabilities induced by the edges} (i.e., not the random walk on the network). 
The corresponding objective function is in fact equivalent to $\underline{\mathcal{L}}\left( \ket{\sigma}; \{\ket{\zeta}_{a}\} \right)$ where we ignore the codewords for the initial module \textit{and} the initial node in each trajectory (consequently, the total length of trajectories $\sum_{a=1}^{M} T_{a}$ is also modified to $\sum_{a=1}^{M} T_{a} - M$). 
A summary table of the {\ACL}s is shown in {\Supplementary} Table 1 ({\Sec}~S1). 

\begin{figure}[t!]
  \centering
  \includegraphics[width= 0.62\columnwidth]{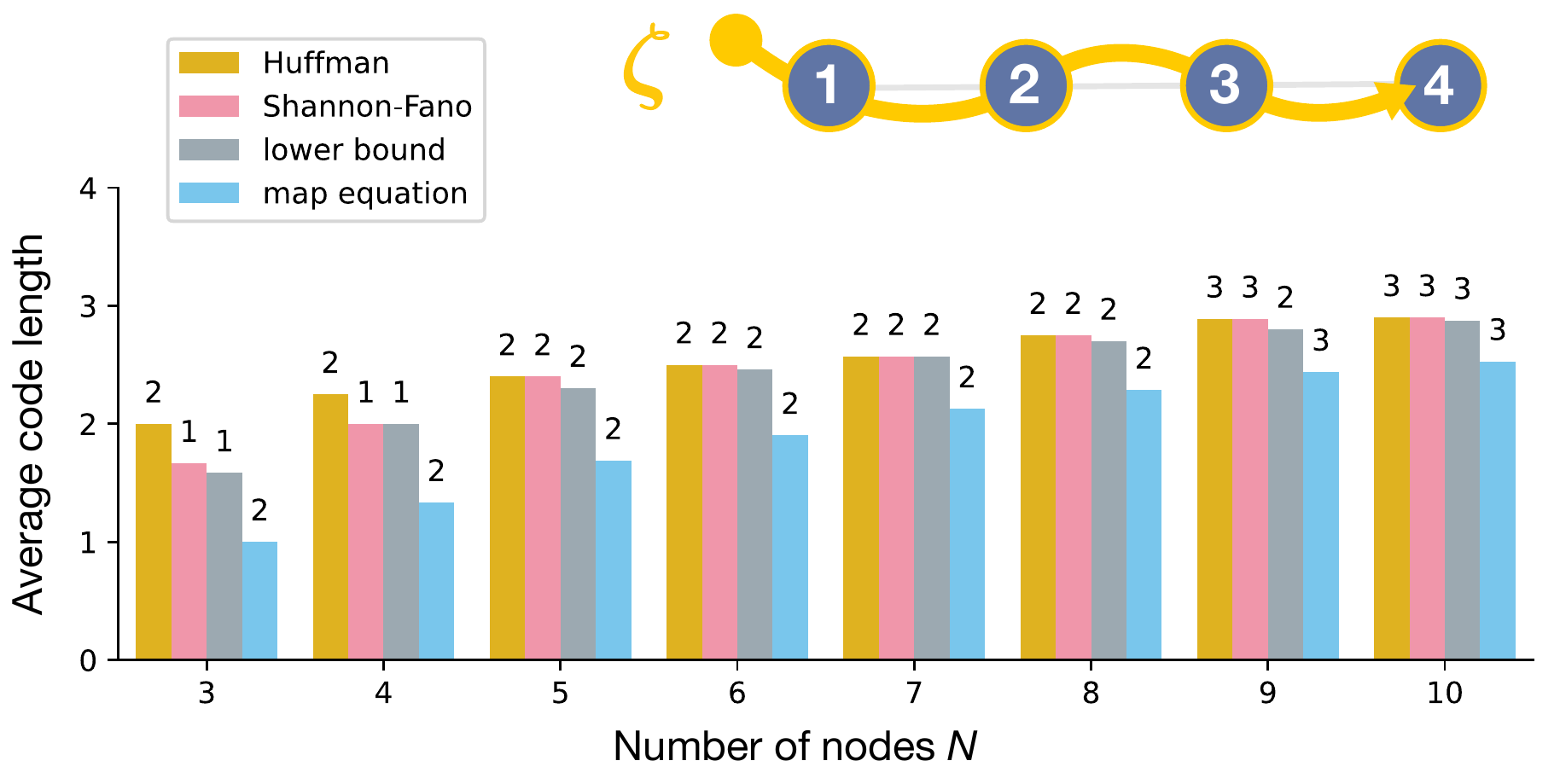}
  \caption{
Average code lengths for a trajectory on a path (illustrated at the top) obtained by minimising $\mathcal{L}\left( \ket{\sigma}; \ket{\zeta} \right)$ (``Huffman'': Huffman coding, ``Shannon-Fano'': Shannon-Fano coding) and $\underline{\mathcal{L}}\left( \ket{\sigma}; \ket{\zeta} \right)$ (``lower bound''). 
The expected average code length $L( \ket{\sigma})$ (``map equation'') based on the set of transition probabilities induced by the edges, i.e., the \texttt{-\,-flow-model rawdir} option, is also shown. 
The number of detected modules in each method is indicated at the top of each bar. 
}
  \label{fig:comparison}
\end{figure}

Before moving on, let us compare how the minimizations of $L( \ket{\sigma})$, $\overline{\mathcal{L}}( \ket{\sigma}; \{\ket{\zeta}_{a}\} )$, and $\underline{\mathcal{L}}\left( \ket{\sigma}; \{\ket{\zeta}_{a}\} \right)$ differ using a simple example. 
We consider a trajectory where a walker visits each node exactly once on a path. 
Figure \ref{fig:comparison} shows the results obtained through the exact minimization of the objective functions. 
It quantifies how the average code lengths approach a common value as $N$ increases, because the contribution from the starting point of the trajectory becomes negligible. 
$\mathcal{L}\left( \ket{\sigma}; \ket{\zeta} \right)$ and $\underline{\mathcal{L}}\left( \ket{\sigma}; \ket{\zeta} \right)$ quickly approach to each other, whereas $L( \ket{\sigma})$ converges relatively slowly, implying that the contribution from the codeword of the initial module can be considerable. 
We also confirmed that the {\stMapEqn} indeed tends to identify a smaller number of modules, and the resulting partitions can vary depending on the coding scheme applied.

The exact minimization of the (expected) {\ACL} is not computationally feasible unless a dataset is extremely small, and thus, we must rely on approximate heuristics in practice. 
The greedy heuristic implemented in Infomap is commonly used for the map equation. 
Therefore, we implemented the optimisation for the {\stMapEqn} as a wrapper of Infomap. 
That is, we first run Infomap as the initial state of the node partition, and then, reduce overfitting by pruning small modules based on $\overline{\mathcal{L}}( \ket{\sigma}; \{\ket{\zeta}_{a}\} )$ or $\underline{\mathcal{L}}\left( \ket{\sigma}; \ket{\zeta} \right)$ as a fine-tuning process; our fine-tuning algorithm is also a greedy heuristic. 
In the following, we refer to this algorithm as Infomap+, and the implementation code is publicly available \cite{GithubURL}. 
Further details of the algorithm are described in {\Methods}. 

\subsection*{Experiments}
This section demonstrates that the {\stMapEqn} prevents overfitting using datasets represented as networks and a real-world dataset as a set of trajectories. 
A network is a special case of trajectory datasets because each directed edge can be regarded as a trajectory with length $T=2$. 
We treat each edge as a pair of directed edges in both directions for an undirected network. 
All networks considered are weakly connected. 

For the network datasets, we can also consider simulated walks on the underlying network as the input trajectories. 
In this setting, we would need to specify the type of simulated walks and choose the values of $T$ and $M$ as hyperparameters. 
Herein, however, we do not consider the simulated walks and treat the edges set directly as the set of trajectories.

\subsubsection*{Network datasets}

\begin{figure}[t!]
  \centering
  \includegraphics[width= 0.7\columnwidth]{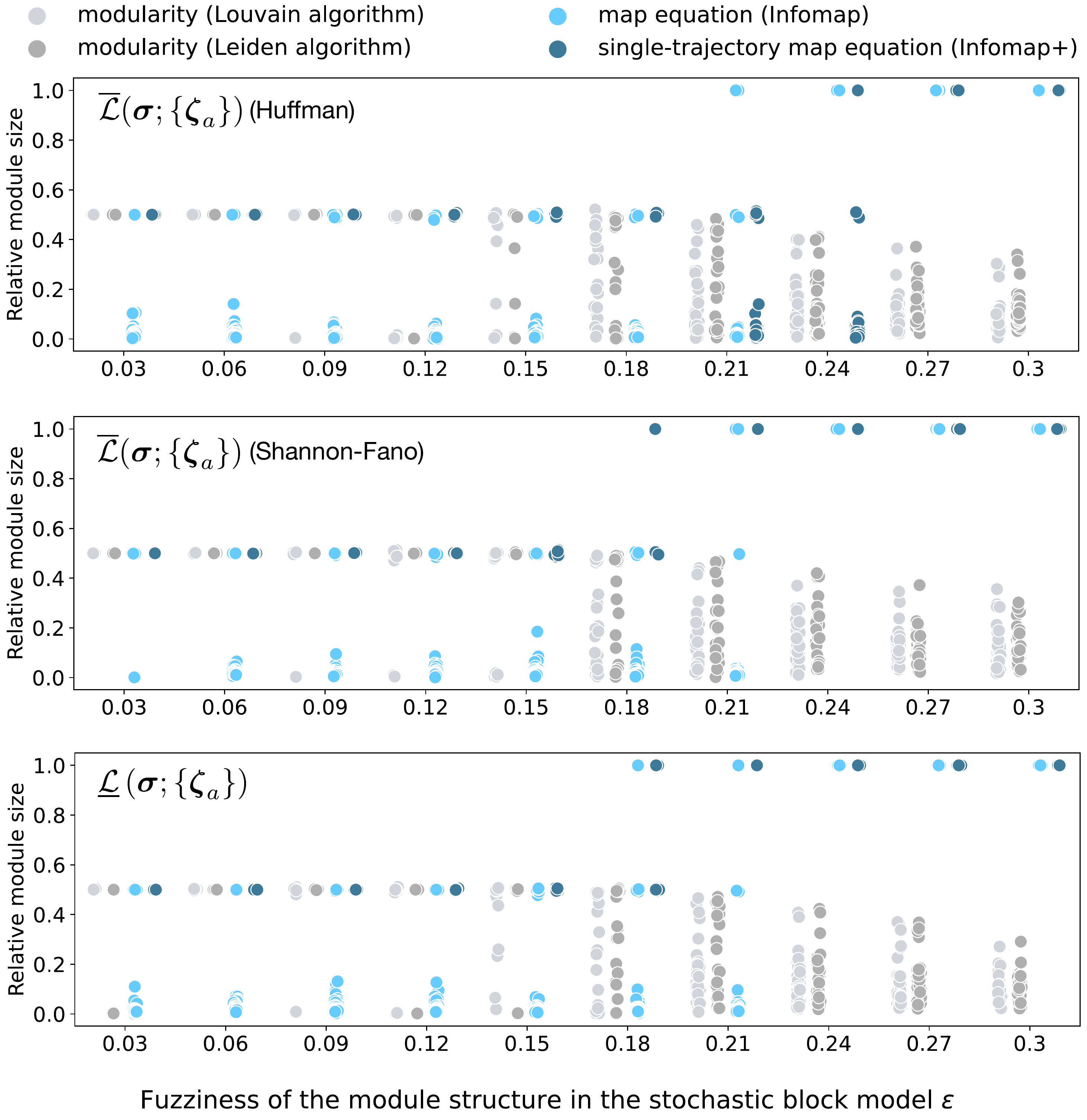}
  \caption{
  Performance of modularity maximization methods (the Louvain and Leiden algorithms), Infomap (\texttt{-\,-two-level}), and algorithms for the {\stMapEqn} based on $\overline{\mathcal{L}}( \ket{\sigma}; \{\ket{\zeta}_{a}\} )$ and $\underline{\mathcal{L}}\left( \ket{\sigma}; \{\ket{\zeta}_{a}\} \right)$ (``Infomap+'') on the symmetric {\SBM} ($N=1,000$, $c=12$). 
  We generated five instances of the {\SBM} for various fuzziness of the module structure $\epsilon$ and plotted the distribution of the resulting relative module sizes. 
  Herein, we set $\lambda = 1$.
The performances of all algorithms change around the algorithmic detectability limit $\epsilon \approx 0.15$, which is distinct from the information-theoretic detectability limit located at $\epsilon = (\sqrt{c}-1)/(\sqrt{c}+1) \simeq 0.55$ \cite{Decelle2011,Mossel2015,Massoulie2014}.
}
  \label{fig:SBMcomparison}
\end{figure}

We first consider synthetic networks that are generated by the stochastic block model ({\SBM}) \cite{holland1983stochastic,WangWong87,AbbeReview2017,Peixoto2017tutorial}, which is a random graph model having a planted (pre-assigned) module structure. 
This is a canonical model that is used for analyses in community detection. 
We particularly consider the so-called symmetric {\SBM} that has two equally-sized planted modules. 
Each pair of nodes in the same planted module is connected with probability $p_{\mathrm{in}}$ and each pair of nodes in different planted modules is connected with probability $p_{\mathrm{out}}$. 
The symmetric {\SBM} is commonly parameterized by the average degree $c$ and the fuzziness of module structure $\epsilon = p_{\mathrm{out}}/p_{\mathrm{in}}$ instead of $p_{\mathrm{in}}$ and $p_{\mathrm{out}}$. 
The detection of planted modules is easier when $\epsilon$ is small because the module structure is clearer. 
Even when $\epsilon < 1$, there exists a critical value of $\epsilon$ above which it becomes impossible to identify the planted module structure better than by chance; this is known as the detectability limit \cite{Decelle2011,Mossel2015,Massoulie2014,AbbeReview2017,MooreReview2017,KawamotoKabashima2019} (in $N \to \infty$). 
For these networks, the {\stMapEqn} cannot be the best method, as the Bayesian inference methods based on the {\SBM} can avoid overfitting at all. 

Figure \ref{fig:SBMcomparison} shows the results of community detection based on the map equation and the {\stMapEqn} applied to the {\SBM}. 
Each point represents the relative module size, which is defined as $\sum_{i=1}^{N} \delta_{\sigma_{i}, \sigma}/N$ for module $\sigma$. 
The results based on modularity maximization (the Louvain \cite{Blondel2008} and Leiden \cite{Traag2019leiden} algorithms) are also shown for comparison. 
The \texttt{-\,-two-level} option in Infomap indicates that it is the method introduced in the original paper for the map equation. 

The Infomap (incorrectly) identifies several small modules even when the module structure is relatively clear, whereas Infomap+ prunes such small modules and identifies the equally-sized modules. 
The network plots in Fig.~\ref{fig:SBM} are the results of the same experiment but with the {\SBM} parameters $N=300$, $c=8$, and $\epsilon=0.1$. 
Although the modularity-based algorithms also identify small modules, Infomap is more prone to overfitting in the region where $\epsilon$ is small. 
This phenomenon can be described by the map equation having a finer resolution limit \cite{KawamotoRosvall2015} compared with that of the modularity \cite{Fortunato2007}, i.e., the map equation can identify smaller modules. 
Note, however, that the analysis of the resolution limit is based on an extreme-case example that has a well-defined module structure; it does not describe the whole behaviour in Fig.~\ref{fig:SBMcomparison}. 
In the region of $\epsilon$ above the detectability limit ($\epsilon \approx 0.15$ \cite{KawamotoKabashima2019}), the modularity-based algorithms subdivide the planted modules into a number of smaller-sized modules. 
This is problematic because a practitioner can hardly realise when the resulting partition is due to overfitting. 
In contrast, most of the map equation-based algorithms do not partition a network in that region, implying that they avoid overfitting. 

Although we showed the relative module sizes obtained by the algorithms, readers might wonder whether the identified modules are actually consistent with the planted ones. 
In {\Supplementary} Fig.~1 ({\Sec}~S2), we confirmed that the inferred and planted module structures are indeed highly consistent when the number of modules is correctly estimated. 
Note also that community detection algorithms generally suffer from overfitting and underfitting more severely when the average degree $c$ is smaller. 
Therefore, all the methods considered here are expected to perform less accurately when $c$ is extremely small. 

The experiments here can be conducted for larger networks. 
In that case, however, some of the plots in Fig.~\ref{fig:SBMcomparison} would be unnecessarily difficult to read because we would have many more points due to ovefitting. 
Moreover, the comparison with the previously known result on the detectability limit may be difficult for larger networks, because it is observed in \cite{KawamotoKabashima2019} that algorithmic detectability limits of greedy algorithms can be size-dependent.

\begin{figure*}[t!]
  \centering
  \includegraphics[width= 0.95\columnwidth]{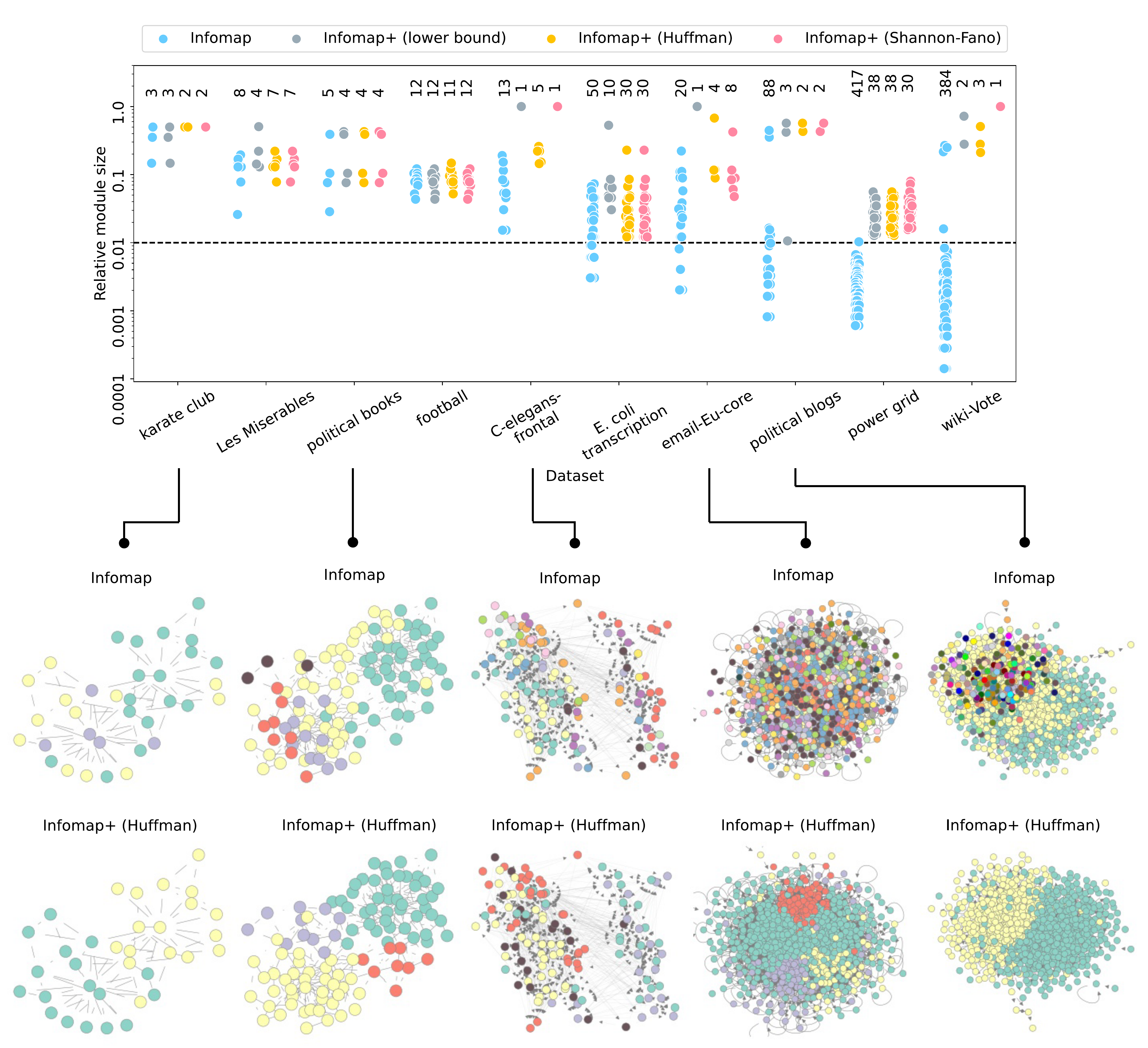}
  \caption{
  Relative module sizes obtained by Infomap and Infomap+ for real-world networks. 
  The number of identified modules is depicted at the top of each result. 
  The dashed line represents $0.01$. 
}
  \label{fig:Realworldcomparison}
\end{figure*}

We then apply the algorithms for the {\stMapEqn} to real-world networks. 
Figure \ref{fig:Realworldcomparison} shows the relative module sizes obtained using Infomap and Infomap+. 
It shows that small modules are pruned, yet larger modules remain identified in most of the cases with Infomap+. 
Although all variants of Infomap+ often provide similar partitions, empirically, the Huffman coding method finds a good balance of module sizes in real-world networks. 
The datasets considered here are often analyzed in the literature on community detection. 
For example, readers can compare the results here with those of Bayesian inference methods reported in \cite{Ghasemian2020,Peixoto2015,Peixoto2017,Kawamoto2017,Kawamoto2018}. 

\begin{figure}[ht!]
  \centering
  \includegraphics[width= 0.67\columnwidth]{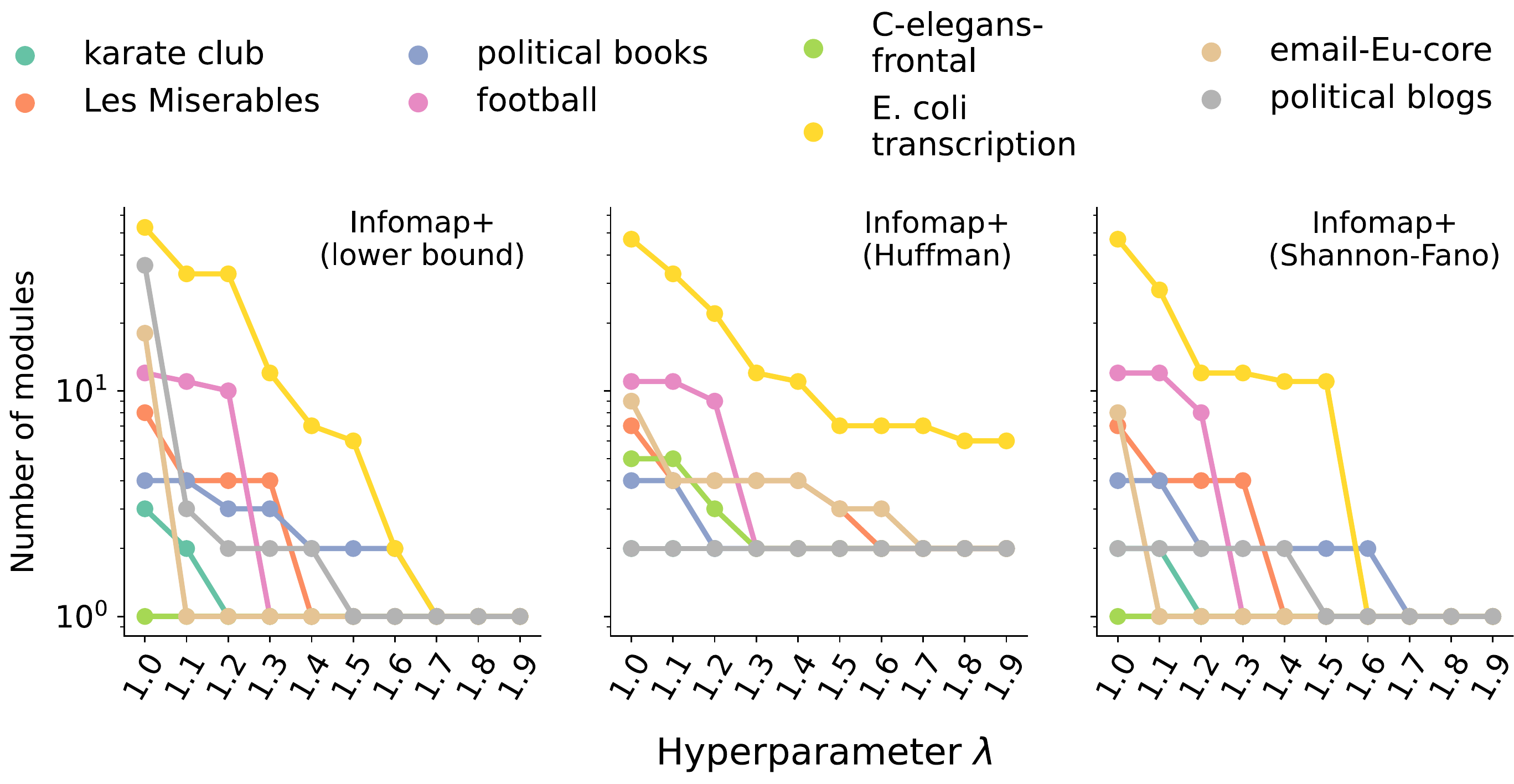}
  \caption{
  Number of modules identified for each value of the hyperparameter $\lambda$ in Infomap+. 
}
  \label{fig:RealworldLambdaDependency}
\end{figure}

In Fig.~\ref{fig:Realworldcomparison}, the value of the hyperparameter $\lambda$, which acts as a resolution parameter, is adjusted for each network so that the size of the smallest module is not less than $\min\{3, N/100\}$ (this adjustment can be performed automatically). 
The selected values of $\lambda$ and the details of experimental settings and datasets are provided in {\Supplementary} Table 2 ({\Sec}~S3). 
We also examined the $\lambda$-dependency in Fig.~\ref{fig:RealworldLambdaDependency}, and we found that the number of modules varies within $1 \le \lambda < 2$ in many datasets; in the Method section, we show that $\lambda=2$ is a practical upper bound according to the resolution limit. 
Note that the threshold $\min\{3, N/100\}$ is only a reference to determine a reasonable value of $\lambda$; when Infomap+ excessively prunes modules, one can directly tune $\lambda$ to resolve the underfitting problem.

\subsubsection*{Bike-sharing dataset: Application to a set of trajectories}
\begin{figure}[t!]
  \centering
  \includegraphics[width= 0.95\columnwidth]{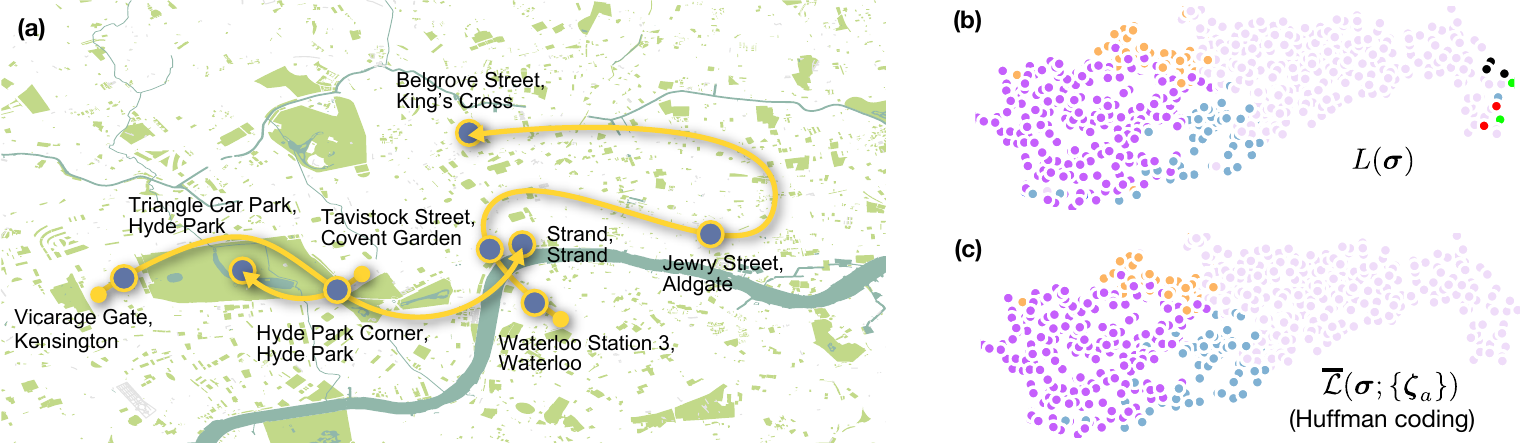}

  \caption{
  Community detection of the bike-sharing dataset. 
  {\bf (a)} Three trajectories in the dataset, where each point (node) represents the location of a bike station. 
  The partitions of the stations are obtained by minimising 
  {\bf (b)} $L( \ket{\sigma})$ (detected $8$ modules) and 
  {\bf (c)} $\overline{\mathcal{L}}( \ket{\sigma}; \{\ket{\zeta}_{a}\} )$ based on the Huffman coding (detected $4$ modules). 
  The stations in the same module are indicated in the same colour.   
}
  \label{fig:LondonBikes}
\end{figure}

Finally, we compare the methods using a dataset of a bike-sharing service in London \cite{MunozMendez2018,BikeShareGithubURL}, which is a dataset consisting of trajectories (sequences of bike stations visited) that individual bikes have travelled in a day (see {\Supplemental} ({\Sec}~S4) for the details of the dataset); thus, $T_{a}$ is the number of stations that bike $a$ has visited. 
Figure \ref{fig:LondonBikes}{\bf (a)} illustrates trajectories of three bikes in the dataset. 
Community detection of the trajectories identifies the area within which a bike is often used. 
Figure \ref{fig:LondonBikes}{\bf (b)} shows the partition obtained by minimising $L( \ket{\sigma})$ using Infomap; here, we constructed a network by decomposing each trajectory into a set of edges between successive pairs of stations. 
As a result, we obtained eight modules; in addition to four large modules, several modules consisting of only a few stations are also identified. 
In Fig.~\ref{fig:LondonBikes}{\bf (c)} which shows the partitions obtained by minimising $\overline{\mathcal{L}}( \ket{\sigma}; \{\ket{\zeta}_{a}\} )$, we no longer observe the small modules. 
Although Fig.~\ref{fig:LondonBikes}{\bf (c)} is of the Huffman coding method ($\lambda=1$), we obtain the same partition with the Shannon-Fano coding method ($\lambda=1$) and with the method minimising $\underline{\mathcal{L}}( \ket{\sigma}; \{\ket{\zeta}_{a}\} )$ ($\lambda=1.8$). 

\section*{Discussion}
This study revisited the formulation of the map equation and shed light on many details hidden in its principle. 
We addressed the fact that the encoding of trajectories is qualitatively distinct from the encoding of the flow on a network and proposed the {\stMapEqn}. 
Importantly, the proposed method can prune small modules and prevent overfitting.

The {\stMapEqn} provides a more balanced community structure compared with the map equation. 
Whereas balanced partitions may not always be desirable, it is often beneficial because we can prune spurious modules due to overfitting as demonstrated in Fig.~\ref{fig:SBM}. 
Furthermore, the analysis in the Method section implies that the {\stMapEqn} is not prone to underfitting compared with the map equation because their resolution limits are almost the same when $\lambda=1$. 

Readers might wonder if the present approach is distinct from other variants of the map equation, such as the one with the Markov-time parameter \cite{Kheirkhahzadeh2016,Schaub2012} and the Bayesian formulation of the map equation \cite{smiljanic2019mapping,smiljanic2021} which is an improved teleportation method \cite{LambiotteRosvall2012}. 
To clarify this point, we also conducted experiments analogous to those described in Experiments section using these methods in {\Supplemental} ({\Sec}~S5). 
In some cases, these methods also exhibit similar partitions as in Figs.~\ref{fig:SBMcomparison} and \ref{fig:Realworldcomparison}. 
However, they are apparently not particularly suitable for pruning small modules because these methods are more sensitive/insensitive to the choices of the hyperparameters that we need to search for the optimal values in a finer scale/wider range, while balanced partitions are often obtained without tuning the hyperparameter $\lambda$ in the {\stMapEqn}. 
We also emphasise that the {\stMapEqn} is not a generalisation of the map equation, but its raw form, and overfitting is avoided using the principle of the map equation itself. 

The bootstrapping method \cite{Rosvall2010} is another approach for avoiding overfitting. 
However, this approach is computationally expensive \cite{smiljanic2021} and a comparison with the present approach is not very clear because the output is a population of partitions. 
A more detailed study of the qualitative and quantitative relationships between the {\stMapEqn} and other variants of the map equation is left for future work. 
Furthermore, because the {\stMapEqn} is a trajectory-based approach, the relationship between the memory-network extension \cite{Rosvall2014,edler2017mapping} of the map equation is another potential research direction because both take a set of trajectories as the input. 

The time complexity of the optimisation algorithm is a major issue in the {\stMapEqn}. 
Whereas the lower bound $\underline{\mathcal{L}}\left( \ket{\sigma}; \{\ket{\zeta}_{a}\} \right)$ can be optimised as efficiently as Infomap, explicit construction of the codebooks is required for the actual average code length $\overline{\mathcal{L}}\left( \ket{\sigma}; \{\ket{\zeta}_{a}\} \right)$. 
Although our implementations of Infomap+ run within a reasonable amount of time for fairly large datasets as demonstrated in {\Supplemental} Fig.~6, an improved implementation is also left for future work.

\section*{Methods}
\subsection*{Optimisation algorithm}
Herein, we explain the implementation details of the greedy heuristic. 
A typical greedy heuristic for community detection, including Infomap, iteratively merges two or more modules that improve the value of an objective function \cite{ClausetNewmanMoore2008,Blondel2008,Traag2019leiden} and equally-sized modules are often preferentially merged \cite{WakitaTsurumi2007}. 
Such an update rule does not effectively compress the {\ACL} at the stage of fine-tuning. 
This is not surprising because the initial partition is located at a local or global minimum of the objective function in the map equation, which may also be a local minimum in the {\stMapEqn}. 
Moreover, there is no reason that equally-sized modules should be preferentially merged. 
Although we typically have a few large and many small modules as the initial partition, it is unlikely that merging those small modules provides better compression of the {\ACL}. 
Therefore, instead, we iteratively merge the smallest module and its most tightly-connected module regardless of the resulting value of the {\ACL} until only one module is left; among the partitions that form with this merging process, we accept the partition that achieves the minimum {\ACL}. 
Given an initial partition, this algorithm is deterministic. 
Although this algorithm is straightforward and the resulting partition may not be the global optimum, an improved compression of the {\ACL} can be achieved by pruning small modules without being trapped into the local minima of the objective function.

When we use the lower bound $\underline{\mathcal{L}}_{\lambda}\left( \ket{\sigma}; \{\ket{\zeta}_{a}\} \right)$ as the objective function, the greedy update can be performed as done in Infomap. 
The expanded form of $\underline{\mathcal{L}}_{\lambda}\left( \ket{\sigma}; \{\ket{\zeta}_{a}\} \right)$ is  
\begin{align}
&\underline{\mathcal{L}}_{\lambda}\left( \ket{\sigma}; \{\ket{\zeta}_{a}\} \right) 
= \lambda \hat{\mathsf{q}}_{\enter} \log \hat{\mathsf{q}}_{\enter} 
-\lambda \sum_{\sigma} \hat{\mathsf{q}}_{\sigma \enter} \log \hat{\mathsf{q}}_{\sigma \enter} 
+ \sum_{\sigma} \hat{\mathsf{p}}^{\sigma}_{\intra} \log \hat{\mathsf{p}}^{\sigma}_{\intra} 
- \sum_{\sigma} \hat{\mathsf{q}}_{\sigma \inter} \log \hat{\mathsf{q}}_{\sigma \inter} 
- \sum_{i=1}^{N} \hat{\mathsf{q}}_{i} \log \hat{\mathsf{q}}_{i}. \label{stMapEqnConcatExpanded}
\end{align}
The last term in {\Eq}~(\ref{stMapEqnConcatExpanded}) is independent of partition. 
Therefore, when we merge two modules, we only need to keep track of changes in $\hat{\mathsf{q}}_{\sigma \enter}$, $\hat{\mathsf{q}}_{\sigma \inter}$, and $\hat{\mathsf{p}}^{\sigma}_{\intra}$, which are defined in {\Eq}~(\ref{qInterIntraProbs}). 
In these quantities, $\sum_{a=1}^{M} \delta_{\sigma^{\prime}, \sigma_{\zeta_{a 0}} }$ is the population of the starting-point nodes in module $\sigma$, 
$\sum_{\sigma} \hat{\mathsf{q}}_{\sigma^{\prime} \sigma}$ and $\sum_{\sigma^{\prime}} \hat{\mathsf{q}}_{\sigma^{\prime} \sigma}$ are the sums of the populations of the transitions across modules in the set of trajectories, and 
$\sum_{i \in \sigma} \hat{\mathsf{q}}_{i}$ is the sum of the node-visiting frequencies in module $\sigma$. 
They are $O(K)$ quantities, such that we can efficiently compute the change in $\underline{\mathcal{L}}_{\lambda}\left( \ket{\sigma}; \{\ket{\zeta}_{a}\} \right)$ when two modules are merged. 

When we use the actual {\ACL} $\overline{\mathcal{L}}_{\lambda}\left( \ket{\sigma}; \{\ket{\zeta}_{a}\} \right)$ as the objective function, the greedy update cannot be computed as efficiently as for $\underline{\mathcal{L}}_{\lambda}\left( \ket{\sigma}; \{\ket{\zeta}_{a}\} \right)$. 
When two modules are merged, we need to reconstruct the intra-module codebook of the target module as well as the inter-module codebook to compute the updated code length. 
The time complexity of constructing a codebook depends on the specific coding scheme applied. 
In {\Supplementary} Fig.~6, we show the running times of Infomap and Infomap+ on the {\SBM}; herein, we used the Infomap API \cite{InfomapAPI} (a C++-based implementation with a Python wrapper) for Infomap and our Python-based implementation \cite{GithubURL} for Infomap+.

\subsection*{Resolution limit}
Readers might consider that the pruning effect implies that the proposed method is prone to underfitting. 
To examine this issue, we derive the resolution limit of the {\stMapEqn} focusing on $\underline{\mathcal{L}}_{\lambda}\left( \ket{\sigma}; \{\ket{\zeta}_{a}\} \right)$ and network datasets. 
The resolution limit is the smallest module size that the method can identify given a network size such as the total number of edges. 

The following analysis shows that, although the method has a relatively coarser-resolution scale compared with the standard or hierarchical map equation, it is still a high-resolution method. 
The analysis also provides a theoretical explanation of some of the empirical results we obtained through the experiments in Experiments section and an implication to the range that the hyperparameter $\lambda$ should take.

\subsubsection*{General form}
We closely follow the derivation in \cite{KawamotoRosvall2015}, which is applied to undirected networks. 
The present resolution limit is for directed networks and the considered objective function is $\underline{\mathcal{L}}_{\lambda}\left( \ket{\sigma}; \{\ket{\zeta}_{a}\} \right)$ instead of $L\left( \ket{\sigma} \right)$. 

We first rewrite the empirical frequencies of the walkers and the objective function in the {\stMapEqn} in terms of network statistics. 
When the input trajectories are the edges in a directed network (i.e., the number of trajectories $M$ is the number of directed edges), we have 
\begin{align}
&\hat{\mathsf{q}}_{\sigma \inter} = \frac{\ell^{\mathrm{out}}_{\sigma}}{2M}, \quad 
\hat{\mathsf{q}}_{\sigma \enter} = \frac{\ell_{\sigma} + \ell^{\mathrm{out}}_{\sigma}}{2M} + \frac{\ell^{\mathrm{in}}_{\sigma}}{2M}, \quad 
\hat{\mathsf{q}}_{\enter} 
= \frac{M+C}{2M}, \quad 
&\hat{\mathsf{p}}^{\sigma}_{\intra} 
= \frac{\ell^{\mathrm{out}}_{\sigma}}{2M} + \frac{2 \ell_{\sigma} + \ell^{\mathrm{out}}_{\sigma} + \ell^{\mathrm{in}}_{\sigma}}{2M}, \quad 
\hat{\mathsf{q}}_{i} = \frac{d^{\mathrm{in}}_{i} + d^{\mathrm{out}}_{i}}{2M},  
\label{NetworkStatFrequencies}
\end{align}
where $\ell_{\sigma}$ is the number of directed edges within module $\sigma$; $\ell^{\mathrm{in}}_{\sigma}$ and $\ell^{\mathrm{out}}_{\sigma}$ are the numbers of in-coming and out-going edges of module $\sigma$, respectively; $d^{\mathrm{in}}_{i}$ and $d^{\mathrm{out}}_{i}$ are the in- and out-degrees of node $i$; 
and $C$ is the cut size of the network, i.e., the total number of directed edges that are crossing different modules. 
Using {\Eq}~(\ref{NetworkStatFrequencies}), the objective function is recast as 
\begin{align}
& \underline{\mathcal{L}}_{\lambda}\left( \ket{\sigma}; \{\ket{\zeta}_{a}\} \right) 
= \frac{1}{2M} \biggl( \lambda (M+C) \log (M+C) + C 
+ \sum_{\sigma} \underline{\mathcal{L}}^{\sigma}_{\lambda}
 + 2M - \sum_{i=1}^{N} \left( d^{\mathrm{in}}_{i} + d^{\mathrm{out}}_{i} \right) \log \left( d^{\mathrm{in}}_{i} + d^{\mathrm{out}}_{i} \right) \biggr),  \label{stMapEqnResolutionLimit1}
\end{align}
where 
\begin{align}
&\underline{\mathcal{L}}^{\sigma}_{\lambda} 
= -\lambda \left( \ell_{\sigma} + \ell^{\mathrm{out}}_{\sigma} + \ell^{\mathrm{in}}_{\sigma} \right) \log \left( \ell_{\sigma} + \ell^{\mathrm{out}}_{\sigma} + \ell^{\mathrm{in}}_{\sigma} \right) 
+ 2 \left( \ell_{\sigma} + \ell^{\mathrm{out}}_{\sigma} + \frac{1}{2}\ell^{\mathrm{in}}_{\sigma} \right) \log \left( \ell_{\sigma} + \ell^{\mathrm{out}}_{\sigma} + \frac{1}{2} \ell^{\mathrm{in}}_{\sigma} \right) 
- \ell^{\mathrm{out}}_{\sigma} \log \ell^{\mathrm{out}}_{\sigma}. \label{stMapEqnResolutionLimit2}
\end{align}

In the resolution-limit analysis, we consider two well-defined modules and derive the condition under which their merging is favoured (i.e., the modules are not resolved) for better optimisation of the objective function. 
Thus, we evaluate the condition such that the difference in the objective function $\Delta \underline{\mathcal{L}}^{\sigma}_{\lambda}$ becomes negative when two modules are merged. 
We denote the labels of two well-defined modules as $A$ and $B$ and the merged module as $AB$. 
We also denote the change in $\sum_{\sigma} \underline{\mathcal{L}}^{\sigma}_{\lambda}$ through the update as $R$, i.e., 
\begin{align}
R = \underline{\mathcal{L}}^{AB}_{\lambda} - \underline{\mathcal{L}}^{A}_{\lambda} - \underline{\mathcal{L}}^{B}_{\lambda}.
\end{align}
Here, $R$ is a local quantity that depends only on the variables within/around modules $A$ and $B$. 
When two well-defined modules are merged, the cut size is decreased by a small $\delta$ ($\delta \ll M + C$). 
The difference in the objective function based on the update is 
\begin{align}
\Delta \underline{\mathcal{L}}_{\lambda}\left( \ket{\sigma}; \{\ket{\zeta}_{a}\} \right) 
&= \frac{1}{2M} \biggl( \lambda (M+C-\delta) \log (M+C-\delta) 
- \lambda (M+C) \log (M+C) - \delta + R \biggr) \notag\\
&\simeq \frac{1}{2M} \biggl( -\delta \lambda \log \left( e (M+C) \right) - \delta + R \biggr), \label{stMapEqnResolutionLimit3}
\end{align}
where $e$ is the basis of the natural logarithm. 
Therefore, the resolution limit is generally expressed as 
\begin{align}
R \lesssim \delta \biggl( 1 + \lambda \log \left( e (M+C) \right) \biggr). \label{stMapEqnResolutionLimit4}
\end{align}
In the map equation, the cut size $C$ is the only global term that is responsible for the resolution limit (see {\Eq}~(11) in \cite{KawamotoRosvall2015}). 
By contrast, the {\stMapEqn} has the total number of directed edges $M$ as another global term in {\Eq}~(\ref{stMapEqnResolutionLimit4}). 
Note, however, that the contribution from $M$ is logarithmic, implying that the {\stMapEqn} is still a high-resolution method. 
Next, we will derive a more explicit scaling.

\subsubsection*{Ring of cliques}
\begin{figure}[t!]
  \centering
  \includegraphics[width= 0.75\columnwidth]{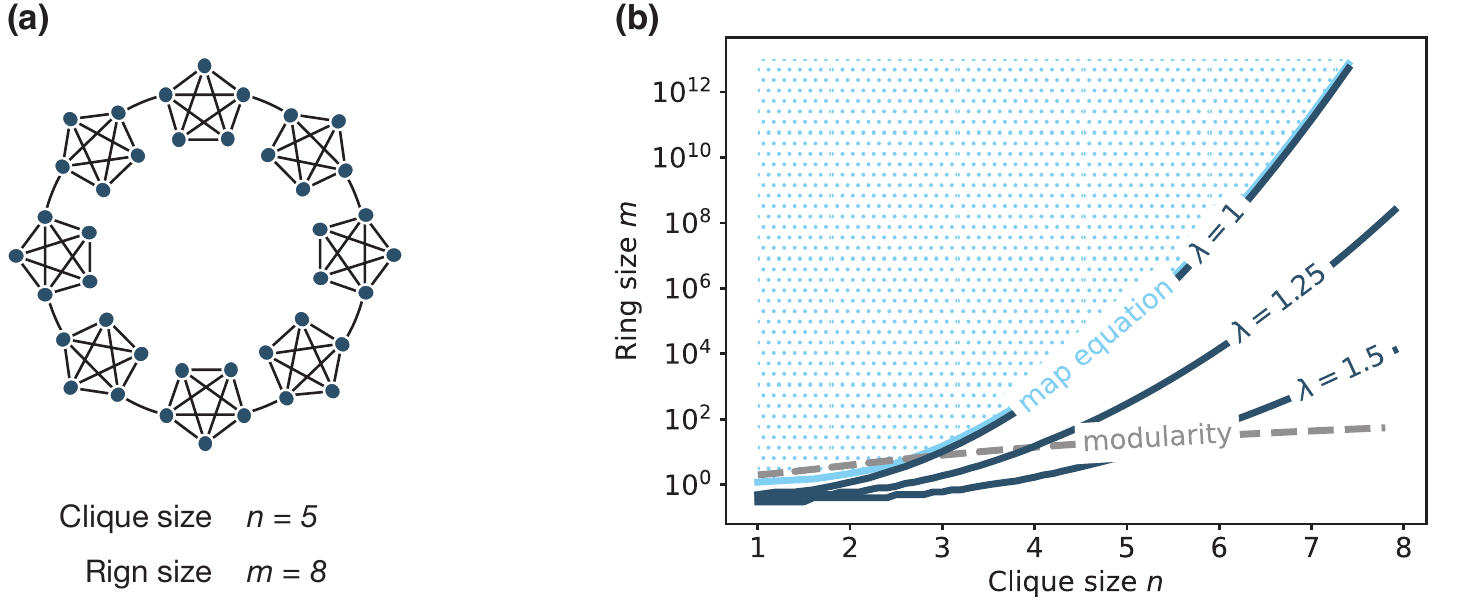}
  \caption{
  	Ring of cliques and its resolution limit. 
	{\bf (a)} Network plot with $n=5$ and $m=8$, and 
	{\bf (b)} the resolution limits of the map equation (light blue line), the {\stMapEqn} with $\lambda=(1, 1.25, 1.5)$ (dark blue lines), and the modularity (dashed grey line). 
	Each line represents the phase boundary above which a clique is not resolved as a module because the network is too large (e.g., the marked region represents the undetectable region in the map equation). 
}
  \label{fig:ResolutionLimit}
\end{figure}

It is common to consider a ``ring of cliques'' in a resolution-limit analysis, as illustrated in Fig.~\ref{fig:ResolutionLimit}{\bf (a)}. 
We consider $m$ cliques (each of which consists of $n$ nodes) and connect each with a single edge to form a ring. 
This is an undirected network. 
Again, we treat each undirected edge as a pair of directed edges in both directions. 
We regard each clique as a module, and using this example, derive the resolution limit in a more explicit form. 

When we merge two of these cliques, the cut size is decreased by 2. 
We denote $\ell_{\sigma} = n (n-1) = \ell$ for an arbitrary module. 
Assuming that $\ell \gg 1$, we have 
\begin{align}
R \simeq -2 (\lambda-2) (\ell + 3) + 2(\lambda-1) \left( \log (e\ell) \right). \label{Rapprox}
\end{align}
Substituting {\Eq}~(\ref{Rapprox}) into {\Eq}~(\ref{stMapEqnResolutionLimit4}), we obtain 
\begin{align}
\frac{\ell^{\lambda-1} 2^{(2-\lambda)(\ell+3)-1}}{e} \lesssim \left( M+C \right)^{\lambda}. \label{stMapEqnResolutionLimit6}
\end{align}
Each clique is resolved as a module unless $(M+C)^{\lambda}$ is larger than the left-hand side of {\Eq}~(\ref{stMapEqnResolutionLimit6}), which is an exponentially growing function with respect to the clique size $\ell$. 

Figure \ref{fig:ResolutionLimit}{\bf (b)} depicts the resolution limits of the {\stMapEqn}, together with those of the map equation \cite{KawamotoRosvall2015} and modularity \cite{Fortunato2007}. 
Although $n$ and $m$ are integers, we treat them as real numbers to highlight the scaling of each resolution limit. 
The resolution limit with $\lambda=1$ is extremely close to that of the map equation. 
Therefore, the {\stMapEqn} is not prone to underfitting compared with the map equation. 

When $\lambda$ is large, modules with a small $n$ are not resolved for any network size. 
However, the limit rapidly disappears as $n$ becomes larger, whereas the resolution limit of the modularity disappears relatively slowly. 
This dependency of the resolution limit partially explains the favourable behaviour of the {\stMapEqn}, i.e., small modules are pruned yet large modules continue to be identified. 
However, as pointed out in the main text, the resolution limit does not describe the full behaviour of the {\stMapEqn}; it is not $\lambda$ that plays a critical role in the method and $\lambda=1$ is often sufficient to avoid overfitting.

In the left-hand side of {\Eq}~(\ref{stMapEqnResolutionLimit6}), the leading coefficient in the exponent becomes negative at $\lambda = 2$. 
In this case, a clique will not be resolved as a module for any network size regardless of its size $n$, i.e., the ability as a community detection method will be completely lost. 
This transition implies that the optimal value of $\lambda$ is usually located within $1 \le \lambda < 2$, which is indeed consistent with our experimental results in Fig.~\ref{fig:RealworldLambdaDependency}.

\begin{acknowledgments}
The author is grateful to Martin Rosvall for inspiring discussions. 
The author also thanks Christopher Bl\"ocker and Teruyoshi Kobayashi for fruitful comments. 
This work was supported by JST ACT-X Grant No. JPMJAX21A8 and JSPS KAKENHI No. 19H01506.
\end{acknowledgments}


%

\clearpage

\setcounter{equation}{0}
\setcounter{figure}{0}
\setcounter{table}{0}
\setcounter{page}{1}
\setcounter{section}{0}

\makeatletter
\renewcommand{\theequation}{S\arabic{equation}}
\renewcommand{\thefigure}{S\arabic{figure}}
\renewcommand{\thesection}{S\arabic{section}}

\renewcommand{\bibnumfmt}[1]{[S#1]}
\renewcommand{\citenumfont}[1]{S#1}

\begin{widetext}

{\Large Supplemental Materials} 

\vspace{.5cm}

{\Large ``Single-trajectory map equation''}\\

{\large Tatsuro Kawamoto}

\vspace{1cm}

\section{Summary of the {\ACL}s}
In Table \ref{table:SymbolTable}, we show a summary table of the average code lengths considered in this study. 

\begin{table*}[h!]
\caption{Symbols and descriptions for the average code lengths considered in the map equation and the {\stMapEqn}.}
\label{table:SymbolTable}
\begin{tabular}{ll}
\hline
Symbol                                                                              & Description                                                                                                                                                                                                                                                                                                                                                                                                                                        \\ \hline
$\mathcal{L}\left( \ket{\zeta}, \ket{\sigma} \right)$                               & Average code length of trajectory $\ket{\zeta}$ with node partition $\ket{\sigma}$                                                                                                                                                                                                                                                                                                                                                                 \\ \hline
$\overline{\mathcal{L}}_{\lambda}\left( \ket{\sigma}; \{\ket{\zeta}_{a}\} \right)$  & \begin{tabular}[c]{@{}l@{}}\tabitem Average code length of trajectories $\{\ket{\zeta}_{a}\}$ with node partition $\ket{\sigma}$ \\ and hyperparameter $\lambda$\\ \tabitem An objective function of the single-trajectory map equation\\ \tabitem $\overline{\mathcal{L}}\left( \ket{\sigma}; \{\ket{\zeta}_{a}\} \right) = \overline{\mathcal{L}}_{\lambda=1}\left( \ket{\sigma}; \{\ket{\zeta}_{a}\} \right)$\end{tabular}                      \\ \hline
$\underline{\mathcal{L}}_{\lambda}\left( \ket{\sigma}; \{\ket{\zeta}_{a}\} \right)$ & \begin{tabular}[c]{@{}l@{}}\tabitem Lower bound of the average code length of trajectories $\{\ket{\zeta}_{a}\}$ \\ with node partition $\ket{\sigma}$ and hyperparameter $\lambda$\\ \tabitem An objective function of the single-trajectory map equation\\ \tabitem $\underline{\mathcal{L}}\left( \ket{\sigma}; \{\ket{\zeta}_{a}\} \right) = \underline{\mathcal{L}}_{\lambda=1}\left( \ket{\sigma}; \{\ket{\zeta}_{a}\} \right)$\end{tabular} \\ \hline
$L\left( \ket{\sigma} \right)$                                                      & \begin{tabular}[c]{@{}l@{}}(\texttt{-\,-two-level} option in Infomap)\\ \tabitem Expected average code length of the random walk with node partition $\ket{\sigma}$\\ \tabitem An objective function of the map equation that is mainly considered \\ in the original paper\end{tabular}                                                                                                                                                           \\ \hline
$L\left( \ket{\sigma} \right)$                                                      & \begin{tabular}[c]{@{}l@{}}(\texttt{-\,-flow-model rawdir} option in Infomap)\\ \tabitem Expected average code length of the flow based on the set of transition \\ probabilities induced by the edges under node partition $\ket{\sigma}$\\ \tabitem An objective function of the map equation that is implemented in Infomap \\ as a variant\end{tabular}                                                                                        \\ \hline
\end{tabular}
\end{table*}

 \clearpage



\section{Accuracy of Infomap+ on the {\SBM}}
For the experiment conducted on the {\SBM} in the main text, we examined whether Infomap+ can correctly estimate the planted number of modules. 
Readers might have doubts whtether the partitions obtained by Infomap+ are consistent with the planted module structure even when the number of modules is accurately estimated.  
To clarify this point, we conducted the same experiment on the {\SBM} and measured the fraction of the correctly classified nodes, which is defined as 
\begin{align}
\frac{1}{N} \max \left\{ \sum_{i=1}^{N} \delta_{\sigma_{i}, \sigma^{\ast}_{i}}, \sum_{i=1}^{N} (1 - \delta_{\sigma_{i}, \sigma^{\ast}_{i}}) \right\}, \label{CorrectFraction}
\end{align}
where $\sigma_{i} \in \{1,2\}$ is the inferred module label and $\sigma^{\ast}_{i} \in \{1,2\}$ is the planted module label. 
Note that $\sum_{i=1}^{N} (1 - \delta_{\sigma_{i}, \sigma^{\ast}_{i}})/N = 1$ indicates that the algorithm perfectly inferred the planted module structure, but with the opposite module label for each node. 
The value of Eq.~(\ref{CorrectFraction}) ranges from $0.5$ to $1$. 
Figure \ref{fig:SBMoverlap} shows that, when Infomap+ correctly estimates the planted number of modules ($K=2$), the fraction of correctly classified nodes is indeed high. 

\begin{figure*}[ht!]
  \centering
  \includegraphics[width= 0.7\columnwidth]{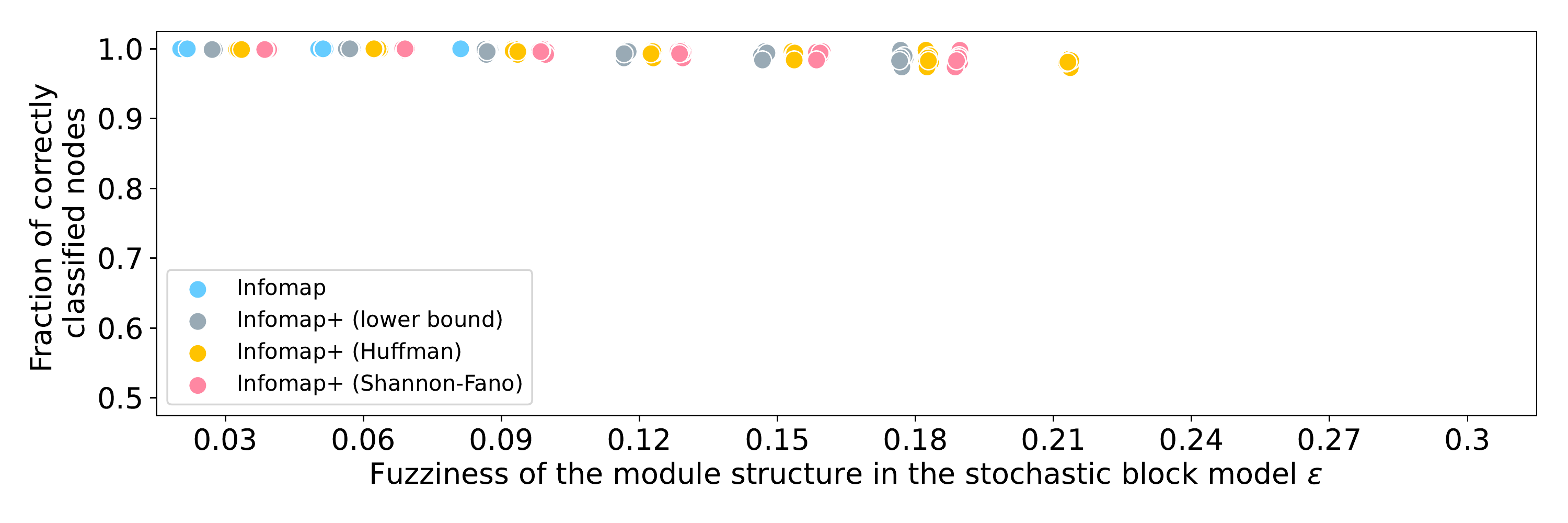}
  \caption{
  Fraction of correctly classified nodes by Infomap and Infomap+ in networks generated using the symmetric {\SBM} ($N=1,000$, $c=12$). 
  We generated 10 instances of the {\SBM} for each fuzziness of the module structure $\epsilon$ and omitted the partitions with $K \ne 2$. 
  Again, we set $\lambda = 1$.
}
  \label{fig:SBMoverlap}
\end{figure*}


\section{Details of the experiments on real-world network datasets}
This section describes the details of the real-world networks analysed in the main text and the settings of the algorithms applied. 
Table~\ref{table:RealworldNetworks} shows the types, number of nodes, and number of edges of the datasets (see the references for the description of each dataset). 
Table~\ref{table:RealworldNetworks} also lists the value of the hyperparameter $\lambda$ used in the {\stMapEqn}. 
Although the Les Miserables network was originally distributed as a weighted network, we converted it to a network with multiple edges because the edge weight represents the number of scene coappearances of characters. 

Recall that a large value of $\lambda$ penalises the generation of new modules. 
Therefore, starting with $\lambda = 1$, we increased the value of $\lambda$ little by little (here, with a step of $0.1$) until the size of the smallest module became sufficiently large. 
As far as we have investigated, $\lambda = 1$ is often already sufficient. 

As the option in Infomap, we used \texttt{-\,-two-level} for the undirected networks and \texttt{-\,-two-level, --directed} for the directed networks (these are the methods introduced in the original paper on the map equation). 
The comparative analysis is fairer and more nontrivial with these options than the experiments with the \texttt{-\,-flow-model rawdir} option because the distinction between the flow-based method and trajectory-based method becomes more prominent. 
We employed the \texttt{-\,-two-level} constraint because the evaluation of multilevel partitioning is beyond the scope of the present study. 

We can obtain a smaller-sized module or a smaller number of modules using Infomap+ than when using Infomap. 
This is because Infomap returns different results with different options (recall that we use \texttt{-\,-flow-model rawdir} for the initialisation in Infomap+) or different runs.

\begin{table*}[t!]
\caption{Real-world network datasets analysed in the main text and the value of the hyperparameter $\lambda$ (for the lower bound, Huffman coding, and Shannon-Fano coding) used in the {\stMapEqn}.}
\label{table:RealworldNetworks}
\begin{tabular}{llllll}
\hline
Dataset name          & Type                     & Nodes & Edges   & References                                       & $\lambda$ \\ \hline
karate club           & undirected, simple graph & 34    & 78      & \cite{karateclub,NewmanNetworkRepo}               & 1, 1, 1         \\
Les Miserables        & undirected, multigraph   & 77    & 254     & \cite{NewmanNetworkRepo}                          & 1, 1, 1         \\
political books       & undirected, simple graph & 105   & 441     & \cite{Newman2006politicalbooks,NewmanNetworkRepo} & 1, 1, 1         \\
football              & undirected, simple graph & 115   & 613     & \cite{NewmanNetworkRepo}                          & 1, 1, 1         \\
C-elegans-frontal     & directed, simple graph   & 131   & 764     & \cite{Celegans,SNAP}                              & 1, 1, 1         \\
E. coli transcription & directed, simple graph   & 328   & 456     & \cite{EcoliTranscription} (v1.1)                  & 1.3, 1.2, 1.1         \\
email-Eu-core         & directed, simple graph   & 986   & 25,552  & \cite{yin2017local,SNAP}                          & 1, 1, 1         \\
political blogs       & directed, simple graph   & 1,222 & 33,428  & \cite{polblogsAdamicGlance,KONECT}                & 1, 1, 1         \\
power grid            & undirected, simple graph & 4,941 & 6,594   & \cite{NewmanNetworkRepo}                          & 1.6, 1.7, 1.6         \\
wiki-Vote             & directed, simple graph   & 7,066 & 103,663 & \cite{wikivoteLeskovec2010,SNAP}                  & 1, 1, 1        \\ \hline
\end{tabular}
\end{table*}

\section{Bike-sharing dataset}
The bike-sharing dataset analysed in the main text was constructed using the dataset distributed through \cite{BikeShareGithubURL}. 
The original dataset consists of riding records of a bike-sharing service in London during 2014; for each use (travel), we can retrieve the starting and ending stations, the time that the bike is used, and the bike ID.  
Because we have a record of bike IDs, we can track the sequence of the stations that an individual bike has visited. 
Although the dataset in \cite{BikeShareGithubURL} is already a subset of larger raw data, we further conducted filtering based on the following criteria: 
\begin{itemize}
\item We consider the uses on July 31, 2014. 
\item We focus on the bikes that visited any of the 50 stations that are used most frequently. 
\item We only consider the series of uses in which the ending station of the previous use coincides with the starting station of the subsequent use. 
\item We exclude the uses in which the starting and ending stations are identical.
\end{itemize}
Consequently, we obtained a total of 5,423 trajectories. 
We used the \texttt{-\,-two-level --directed} option to run Infomap.

\section{Comparison with other variants of the map equation}

\subsection{Performance of Infomap with the Markov-time parameter}
Figures \ref{fig:InfomapMarkovTimeSBM} and \ref{fig:InfomapMarkovTimeRealworld} show the performance of Infomap with the Markov-time parameter (which we refer to as $\tau$) on the {\SBM} and real-world networks, corresponding to the experiments performed in the main text. 
Note that there is no nontrivial default value of $\tau$; $\tau=1$ corresponds to the standard Infomap, and we must choose a value $\tau>1$ to obtain a result distinct from that without the Markov-time parameter. 
Therefore, we raised $\tau$ from $1$ little by little (with a step of $0.1$) until the size of the smallest module was not less than $\max\{ 3, N/100 \}$; when this condition was not satisfied, we set $\tau = 100$. 

In some cases, this approach also provides a partition similar to that based on the {\stMapEqn}. 
However, because the Markov-time parameter approach globally modifies the resolution scale of partitions, large modules are often merged together as well, at the expense of pruning small modules. 
For example, in the experiment conducted on the {\SBM}, the performance is considerably deteriorated in terms of the detectability limit when the Markov-time parameter is large; that is, the planted modules are merged even when the fuzziness parameter $\epsilon$ is small. 
Note also that the range of $\tau$ we need to sweep is relatively large. 
By contrast, in the {\stMapEqn}, $\lambda=1$ or a slightly larger value is often sufficient to avoid overfitting. 

\begin{figure}[h!]
  \centering
  \includegraphics[width= 0.7\columnwidth]{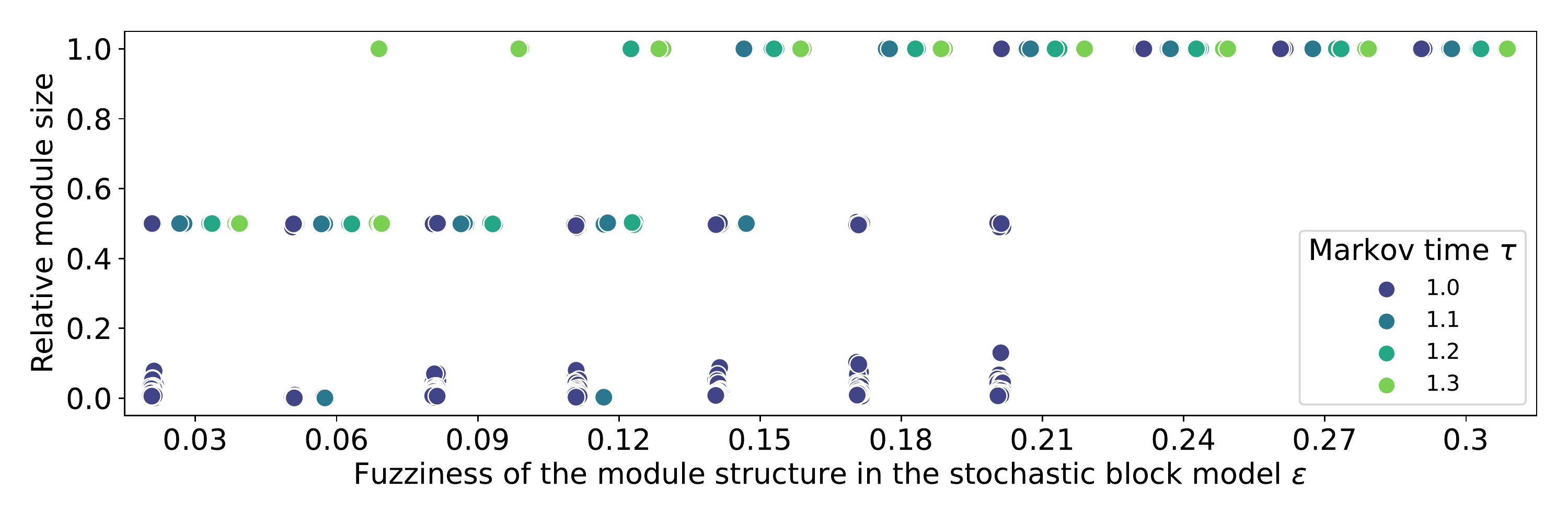}
  \caption{
  Performance of Infomap with different values of the Markov-time parameter on the symmetric {\SBM} ($N=1,000$, $c=12$). 
  We generated five instances of the {\SBM} for each fuzziness of the module structure $\epsilon$ and plotted the distribution of the resulting relative module sizes. 
}
  \label{fig:InfomapMarkovTimeSBM}
\end{figure}

\begin{figure}[h!]
  \centering
  \includegraphics[width= 0.8\columnwidth]{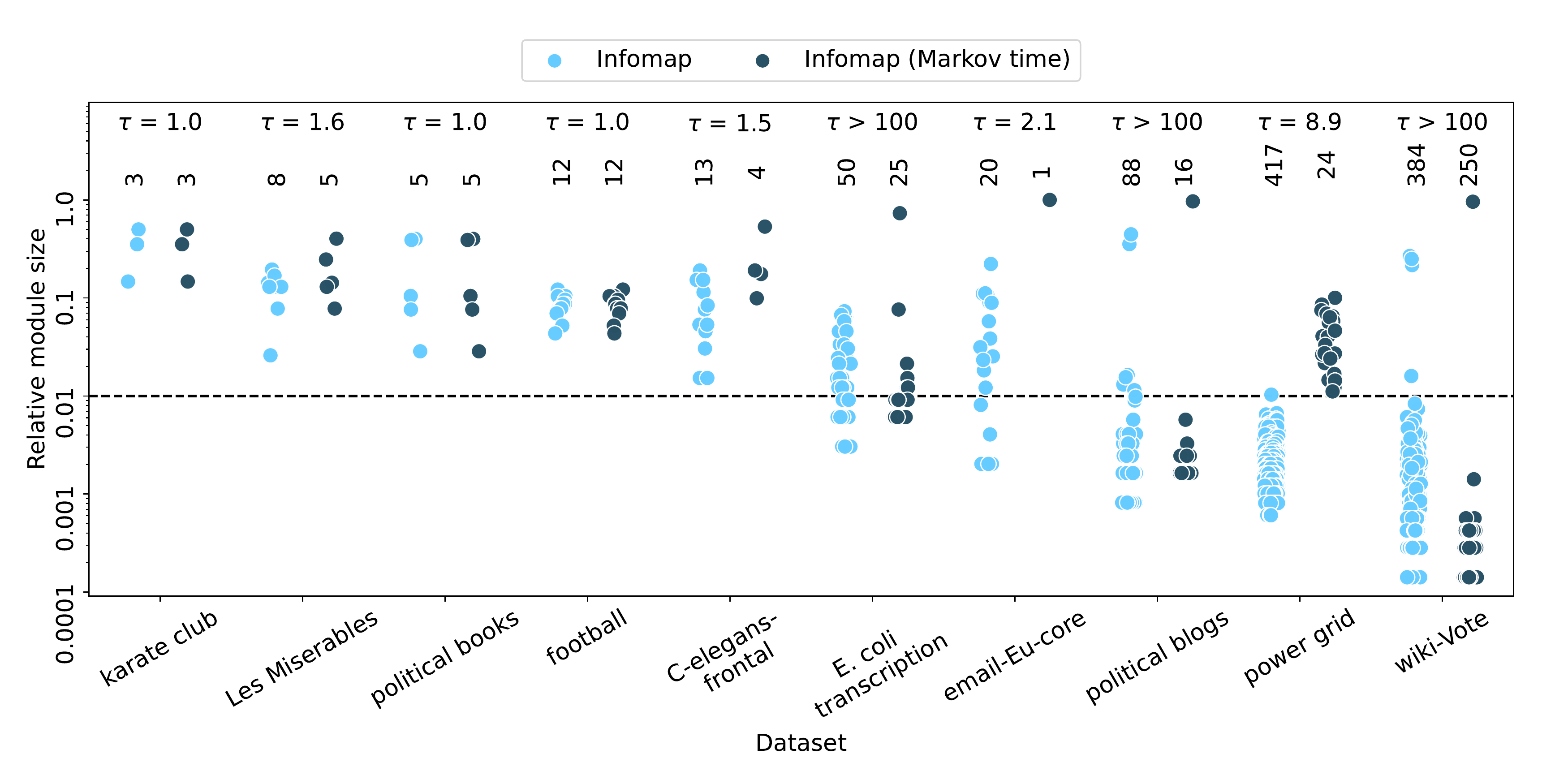}
  \caption{
  Relative module sizes obtained by Infomap (without the Markov-time parameter) and Infomap with different values of the Markov-time parameters on the real-world networks considered in the main text. 
  The number of identified modules and the value of the Markov-time parameter $\tau$ are depicted at the top of each result.  
  We selected the minimum value of $\tau$ such that the smallest module is not less than $\max\{ 3, N/100 \}$ (the dashed line represents $0.01$); otherwise, we set $\tau = 100$ (and denote ``$\tau > 100$''). 
}
  \label{fig:InfomapMarkovTimeRealworld}
\end{figure}

\subsection{Performance of the Bayesian Infomap}
Figures \ref{fig:BayesInfomapSBM} and \ref{fig:BayesInfomapRealworld} show the performance of the Bayesian Infomap  \cite{smiljanic2019mapping,smiljanic2021} on the SBM and real-world networks, corresponding to the experiments performed in the main text. 
The Bayesian Infomap has a hyperparameter $\tilde{\lambda}$ that specifies the strength of the prior distribution based on a random network. 
The default value is $\tilde{\lambda} = (\ln N)/N$; in Infomap, the parameter \texttt{regularisation\_strength} controls the coefficient $a$ in $\tilde{\lambda} = a (\ln N)/N$. 

Overall, the Bayesian Infomap is highly sensitive to the choice of $\tilde{\lambda}$. 
As observed in Fig.~\ref{fig:BayesInfomapSBM}, in many cases, the Bayesian Infomap either leaves many small modules or identifies the whole network as a module. 
The same tendency was observed for the real-world networks, as shown in Fig.~\ref{fig:BayesInfomapRealworld}. 
Similar to the experiment in the main text, we increased the value of $\tilde{\lambda}$ from zero (with a step of $0.1$) until the size of the smallest module was not less than $\max\{ 3, N/100 \}$. 
As a result, the Bayesian Infomap did not identify nontrivial modules for the large networks. 
We also conducted a version in which the threshold of the smallest module was $\max\{ 3, N/1,000 \}$. 
However, the same number of modules was obtained for each dataset. 

In summary, although the Bayesian Infomap also aims to avoid overfitting, its performance is distinct from that of the {\stMapEqn}. 
For the datasets we have investigated, it was not easy to prune small modules while continuing to identify large modules. 
However, it should also be noted that the Bayesian Infomap is a highly flexible method that the performance can be improved by tuning the prior distribution more carefully.  
 
\begin{figure}[h!]
  \centering
  \includegraphics[width= 0.7\columnwidth]{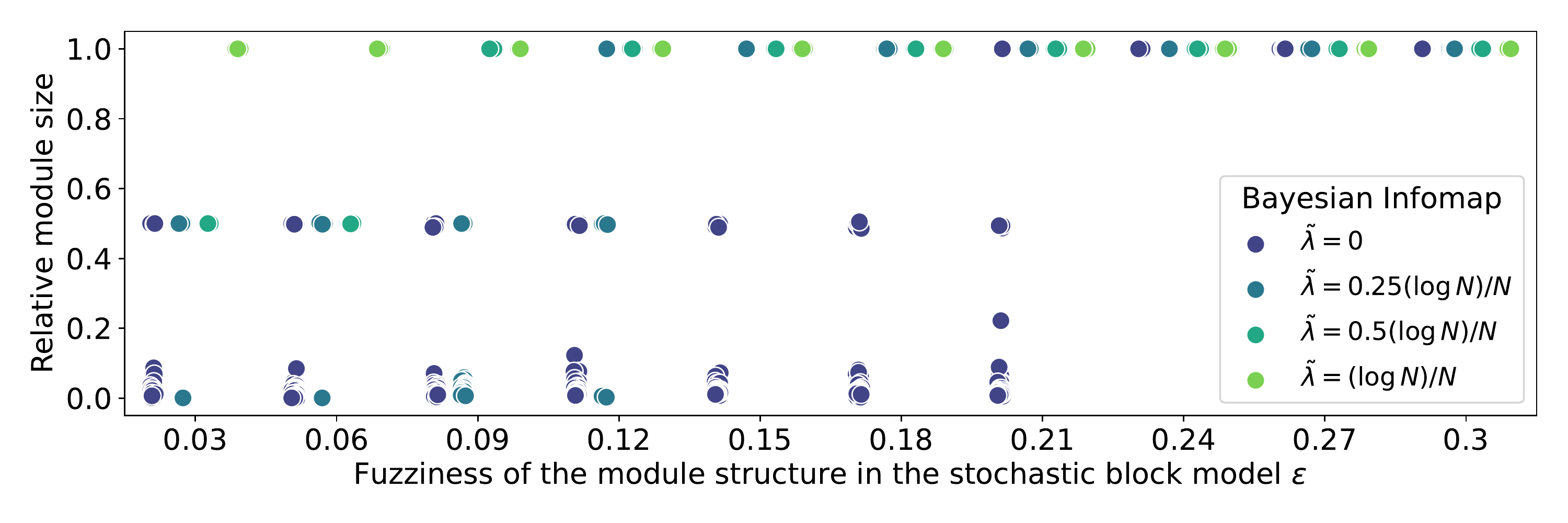}
  \caption{
  Performance of the Bayesian Infomap with different values of the prior parameter $\tilde{\lambda}$ on the symmetric {\SBM} ($N=1,000$, $c=12$). 
  We generated five instances of the {\SBM} for each fuzziness of the module structure $\epsilon$ and plotted the distribution of the resulting relative module sizes. 
}
  \label{fig:BayesInfomapSBM}
\end{figure}

\begin{figure}[h!]
  \centering
  \includegraphics[width= 0.8\columnwidth]{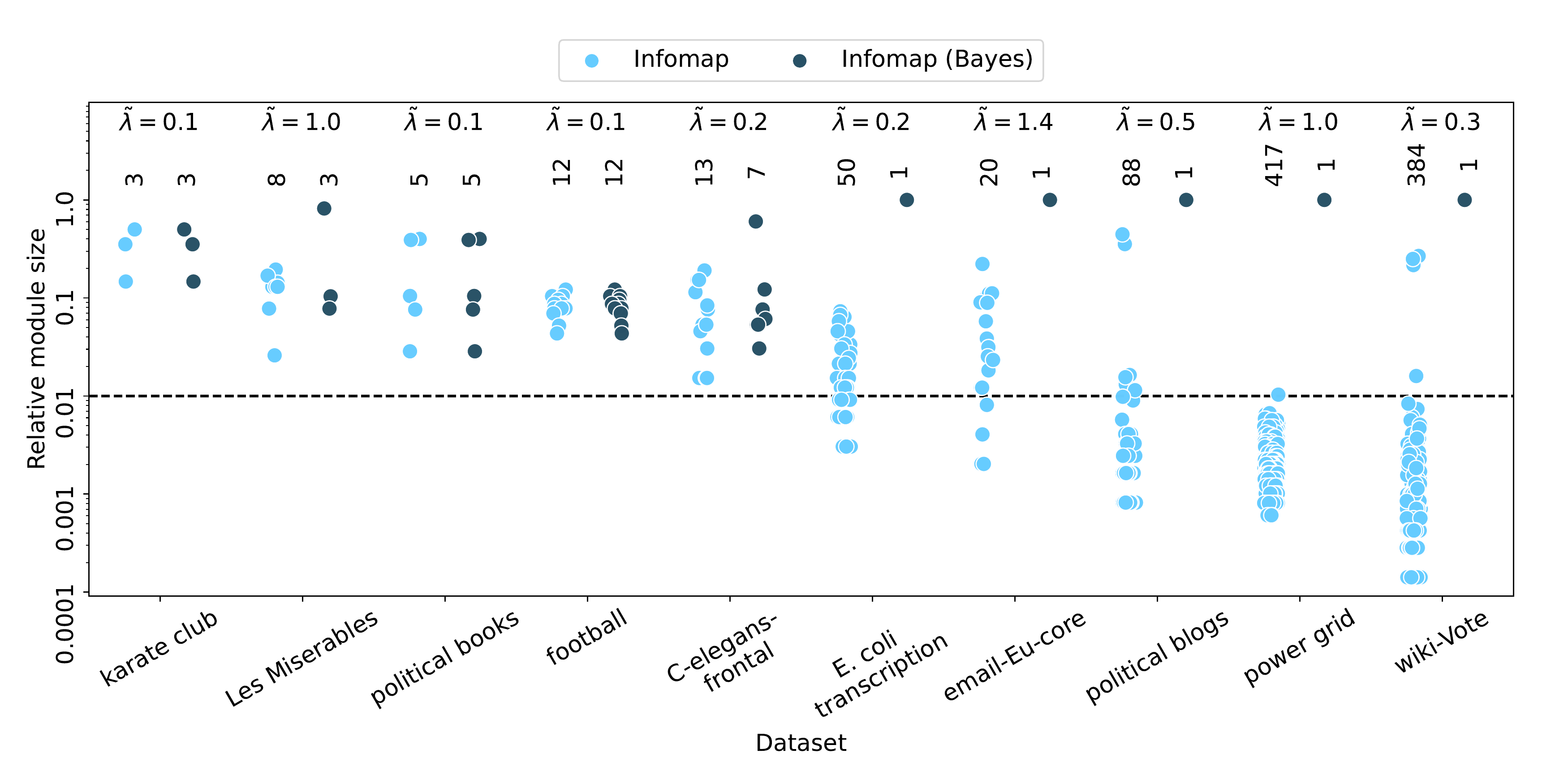}
  \caption{
  Relative module sizes obtained by Infomap (without the Bayesian regularisation) and the Bayesian Infomap with different values of the prior parameter for the real-world networks considered in the main text. 
  The number of identified modules and the value of the prior parameter $\tilde{\lambda}$ are depicted at the top of each result.  
  We selected the minimum value of $\tilde{\lambda}$ such that the smallest module is not less than $\max\{ 3, N/100 \}$ (the dashed line represents $0.01$). 
}
  \label{fig:BayesInfomapRealworld}
\end{figure}

\begin{figure}[t!]
  \centering
  \includegraphics[width= 0.5\columnwidth]{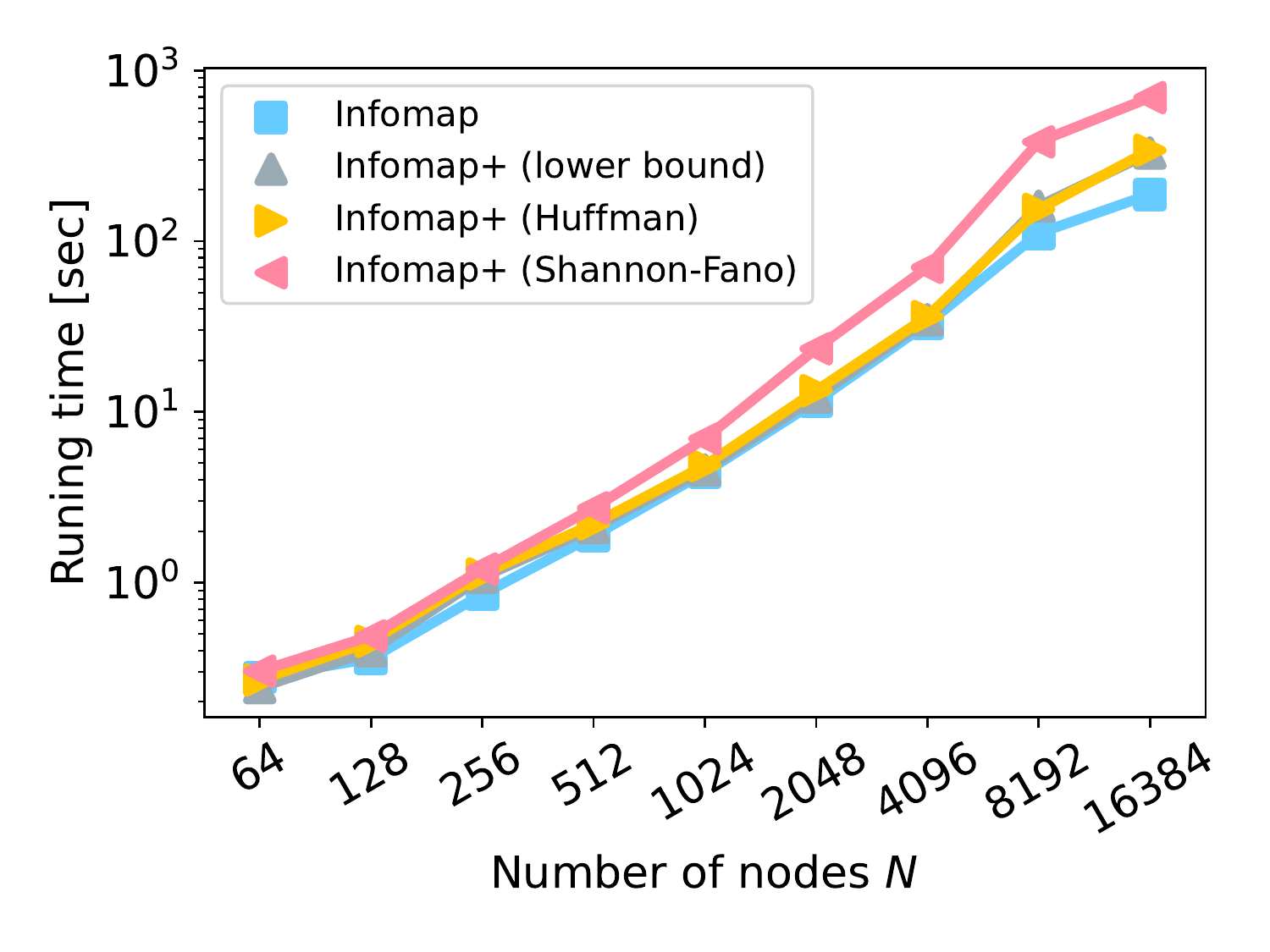}
  \caption{
  Running times of Infomap (\texttt{-\,-two-level}) and Infomap+ on the {\SBM}. 
  Each point represents the mean running time for five network instances generated from the {\SBM} with eight equally-sized planted modules ($c=12$, $\epsilon=0.1$). 
  The running time of each algorithm grows polynomially with $N$. 
}
  \label{fig:RunningTime}
\end{figure}

\end{widetext}


%

\end{document}